\documentclass[useAMS,usenatbib]{mnras}

\usepackage{times}

\usepackage{natbib}
\usepackage{amssymb,amsmath}
\usepackage{graphicx,color}
\usepackage{hyperref}
\usepackage[usenames,dvipsnames]{xcolor}
\usepackage[OT2,T1]{fontenc}
\usepackage[draft]{todonotes}
\newcommand\revy[1]{#1} 
\newcommand\revsw[1]{#1}
\newcommand\revfix[1]{#1}
\newcommand\revall[1]{#1} 
\newcommand\revfinal[1]{#1}
\newcommand\refresponse[1]{#1}

\newcommand{\hkpc}{$h^{-1}$kpc}
\newcommand{\hmpc}{$h^{-1}$Mpc}
\newcommand{\beq}{\begin{equation}}
\newcommand{\eeq}{\end{equation}}

\title[Central galaxy halo masses]{Strong bimodality in the host halo mass of central galaxies from
  galaxy-galaxy lensing}

\author[Mandelbaum et~al.]{Rachel Mandelbaum$^1$\thanks{\tt rmandelb@andrew.cmu.edu},
Wenting Wang$^2$, Ying Zu$^1$, Simon White$^3$,
\newauthor
Bruno Henriques$^{3,4}$, Surhud More$^5$
\\$^1$McWilliams Center for Cosmology, Department of Physics, 
Carnegie Mellon University, Pittsburgh, PA 15213, USA
\\$^2$Institute for Computational Cosmology, University of Durham, South Road, Durham, DH1 3LE, UK
\\$^3$Max Planck Institut fur Astrophysik, Karl-Schwarzschild-Str.\ 1, 85741 Garching b. M{\"u}nchen, Germany
\\$^4$Institute for Astronomy, Department of Physics, ETH Zurich, 8093 Zurich, Switzerland
\\$^5$Kavli Institute for the Physics and Mathematics of the Universe (WPI), UTIAS, The University of Tokyo, Chiba, 277-8583, Japan
}

\begin{document}

\date{\today}

\maketitle

\begin{abstract}
  \revall{
We use galaxy-galaxy lensing to study the dark matter
halos surrounding a sample of Locally Brightest Galaxies (LBGs)
selected from the Sloan Digital Sky Survey. We measure mean halo mass
as a function of the stellar mass and colour of the central galaxy.
Mock catalogues constructed from semi-analytic galaxy formation
simulations demonstrate that most LBGs are the central objects of
their halos, greatly reducing interpretation uncertainties due to
satellite contributions to the lensing signal. Over the full stellar
mass range, \refresponse{$10.3 < \log [M_*/M_\odot] < 11.6$}, we find that passive
central galaxies have halos that are at least twice as massive as
those of star-forming objects of the same stellar mass. The
significance of this effect exceeds $3\sigma$ for \refresponse{$\log
[M_*/M_\odot] > 10.7$}.  \revfinal{Tests using the mock catalogues and on the data
themselves clarify the effects of LBG selection and show that it
cannot artificially induce a systematic dependence of halo mass on LBG
colour. The bimodality in halo mass at fixed stellar mass is reproduced by the astrophysical
model underlying our mock catalogue, but the sign of the effect is inconsistent with
recent, nearly parameter-free age-matching models.  The sign and magnitude of the effect can, however, be reproduced by halo occupation distribution
models  with a simple (few-parameter) prescription for type-dependence.}
}
\end{abstract}

\begin{keywords}
cosmology: observations --- galaxies: haloes --- gravitational lensing: weak --- galaxies: stellar
content
\end{keywords}

%
%
\section{Introduction}

\revsw{The relationship between the stellar mass
in galaxies and the mass of their underlying dark matter halos is a fundamental ingredient in any
theory of galaxy formation and evolution.  This relationship has been
inferred using multiple observational techniques, each with different interpretation difficulties.
These include abundance matching
\citep[e.g.,][]{2010ApJ...717..379B,2010MNRAS.404.1111G,2010ApJ...710..903M}; galaxy clustering
\citep[e.g.,][]{2005ApJ...630....1Z,2011ApJ...728..126W,2011ApJ...736...59Z,2012A&A...542A...5C,2013MNRAS.429...98P}; and
satellite kinematics \citep[e.g.,][]{2007ApJ...654..153C,2011MNRAS.410..210M,2013MNRAS.428.2407W}. 
 Galaxy-galaxy lensing
\citep[e.g.,][]{2004ApJ...606...67H,2006MNRAS.371L..60H,2006MNRAS.368..715M,2014MNRAS.437.2111V,2015MNRAS.446.1356H,2015MNRAS.447..298H,2015ApJ...806....1M,2015A&A...579A..26V}
exploits the weak shape distortions of distant galaxies due to the mass in nearby foreground
galaxies, and therefore has the unique advantage of being directly sensitive to dark matter halo masses.  However,
as for galaxy clustering, the interpretation of the results can be complicated due to the different
lensing profiles of central and satellite galaxies (the latter of which include a contribution from
their host dark matter halo); this is also the case when combining lensing with other 
measurements
\citep[e.g.,][]{2009MNRAS.394..929C,2012ApJ...744..159L,2013MNRAS.430..767C,2015MNRAS.449.1352C,2015ApJ...806....2M,2015MNRAS.454.1161Z}. }

Galaxies are often classified as passive or active\footnote{Throughout this work, we use
  ``passive'' vs.\ ''active'' to indicate galaxy types, and ``red'' vs.\ ``blue'' to indicate rest-frame
color-selected samples (our observational proxies for passive vs.\ active galaxies).}, where passive
galaxies typically have a 
red rest-frame color indicative of a paucity of recent star formation, and active galaxies have
bluer colors indicating recent and/or ongoing star formation.  They also exhibit morphological
differences, and variations in their distributions with respect to the local density (with
passive galaxies more prevalent in high-density environments).  It is natural to ask 
whether these two classes of galaxies, which apparently have rather different star formation
histories, also have different relationships between stellar and halo mass.  Unfortunately, the
answer to this question is observationally 
unclear, with results pointing to differences between passive and active galaxies 
\revsw{\citep{2005ApJ...635...73H,2006MNRAS.368..715M,2011MNRAS.410..210M,2012MNRAS.424.2574W,2013ApJ...778...93T,2014MNRAS.437.2111V,2015MNRAS.447..298H,2015MNRAS.446..521S}},
but often with quite large \refresponse{statistical or systematic} uncertainties that can make it difficult to draw a definitive conclusion,
especially at lower mass.  

The relationship between stellar and halo mass can also be explored using \refresponse{simulation
and/or galaxy modeling techniques}.  For example, semi-analytic models (SAMs) applied to subhalo
catalogs from $N$-body simulations (as early as \citealt{1997MNRAS.286..795K}; see
\citealt{2011MNRAS.413..101G,2012NewA...17..175B} for more recent examples) and hydrodynamic
simulations \revsw{\citep[e.g.,][]{2014Natur.509..177V,2015MNRAS.450.1349K,2015MNRAS.446..521S}} predict a potentially quite
complicated and physically-based relation between the galaxies and their dark matter halos,
including a properly self-consistent treatment of the time evolution of this relationship.  In both
cases, there are parameters that must be tuned to match observations.  For SAMs, these are explicit 
parameters of the semi-analytic model, while for hydrodynamic simulations, they model 
subgrid physics unresolved by the simulation, and again must be tuned 
by hand.  Both methods are usually calibrated to  match a subset of the interesting properties of galaxies
in real data (e.g., mass functions), and then tested against different properties (e.g., clustering
or evolution).  In contrast, the age-matching technique
\citep{2013MNRAS.435.1313H,2014MNRAS.444..729H} can also be
applied to catalogs from an $N$-body simulation, and assumes (rather than predicting) a particular relationship
between the masses and ages of galaxies and those of their halos in order to populate the halos with galaxies.

The various theoretical \refresponse{frameworks} mentioned above take different approaches to the issue of galaxy
type-dependence. Observational guidance on the color-dependence of the relation between stellar and
halo mass using a simple and unambiguous measurement with minimal modeling assumptions may
therefore be used to distinguish between these models.  In this paper, we use a dataset consisting
of Locally Brightest Galaxies (LBGs; \citealt{2013A&A...557A..52P} and
\citealt{2015MNRAS.449.3806A}) in the Sloan Digital Sky Survey (SDSS) as a sample of central
galaxies, classify them as passive or active based on rest-frame colors, and use weak gravitational
lensing to estimate their average halo mass as a function of their stellar masses.  We also present
evidence that this LBG dataset is a relatively fair sample of central galaxies, and use mock
catalogs to correct for biases resulting from the selection method and our way of interpreting the
weak lensing measurements in terms of a halo mass.  \revsw{The use of LBGs thus reduces sensitivity
  to the separation between  central and satellite contributions, which is the primary
  source of model-dependence when interpreting galaxy-galaxy lensing measurements, and we are left
  with a simple and direct measurement of halo masses.}
\revall{\cite{wenting-tmp}} carry out
a thorough exploration of how SAM-based catalogs capture the observed lensing properties of these LBGs,
considering multiple cosmological models and SAMs; in this work, we use the mock catalog that
{\em best} reproduces the observed properties of LBGs to interpret our measurements. 

The outline of this paper is as follows.  In Sec.~\ref{sec:data}, we describe the dataset used for
the measurements, and in Sec.~\ref{sec:method} we describe the measurements made and how they are
interpreted.  Sec.~\ref{sec:mocks} includes a description of the mock catalogs used to validate the
measurement method, and the result of that validation.  We present our results and compare with
previous work in Sec.~\ref{sec:results}, and discuss the implications for galaxy formation and
evolution in Sec.~\ref{sec:discussion}.  

All measurements made in this paper use a flat $\Lambda$CDM model with $\Omega_m=0.315$ in
accordance with the 2013 Planck results in \cite{2014A&A...571A..16P}, and stellar mass estimates
are calculated in $M_\odot$ using $h=0.673$ as in that work.  This choice was made to conform with
the cosmological parameters in the mock galaxy catalog used to validate our measurement procedure.

\section{Data}\label{sec:data}

All data used in this paper came 
from the Sloan Digital Sky Survey (SDSS) \revsw{I/II}.   The SDSS \revsw{I}
\citep{2000AJ....120.1579Y} \revsw{and II surveys} imaged roughly $\pi$ steradians
of the sky, and followed up approximately one million of the detected
objects spectroscopically \citep{2001AJ....122.2267E,
  2002AJ....123.2945R,2002AJ....124.1810S}. The imaging was carried
out by drift-scanning the sky in photometric conditions
\citep{2001AJ....122.2129H, 2004AN....325..583I}, in five bands
($ugriz$) \citep{1996AJ....111.1748F, 2002AJ....123.2121S} using a
specially-designed wide-field camera
\citep{1998AJ....116.3040G}. These imaging data were used to create
the catalogues that we use in this paper.  All of
the data were processed by completely automated pipelines that detect
and measure photometric properties of objects, and astrometrically
calibrate the data \citep{2001ASPC..238..269L,
  2003AJ....125.1559P,2006AN....327..821T}. The SDSS-I/II imaging
surveys were completed with a seventh data release
\citep{2009ApJS..182..543A}, from which the LBG sample described in Sec.~\ref{subsec:lbg}
originates.  The source catalog described in Sec.~\ref{subsec:source} uses an 
improved data reduction pipeline that was part of the eighth data
release, from SDSS-III \citep{2011ApJS..193...29A}; and an improved
photometric calibration \citep[`ubercalibration',][]{2008ApJ...674.1217P}.

\subsection{LBG sample}\label{subsec:lbg}

The LBG sample used in this work was initially defined by \cite{2013A&A...557A..52P} and
\cite{2015MNRAS.449.3806A}, so we closely follow their methodology with only minor modifications
explained below.  They selected the LBGs from the flux-limited SDSS Main galaxy sample, using the
Value Added Galaxy Catalog (VAGC; \citealt{2005AJ....129.2562B}) \revall{\texttt{all0} sample with a
constant flux limit of $r<17.7$}, including absolute magnitudes and stellar mass
estimates using {\sc kcorrect} \citep{2007AJ....133..734B}, which fits stellar population synthesis models to the
five-band photometry assuming a \cite{2003PASP..115..763C} stellar initial mass function (IMF).  This catalog was
based on the seventh data release of the SDSS \citep{2009ApJS..182..543A}.

LBGs were selected in \cite{2013A&A...557A..52P} and \cite{2015MNRAS.449.3806A} by defining
cylinders of radius 1~Mpc, extending $\pm 1000$~km$/$s in the redshift direction, and requiring that
\revsw{each LBG be the brightest galaxy in its cylinder according to the $r$-band absolute magnitude.  In
addition, a photometric redshift catalog \citep{2009MNRAS.396.2379C} was used to eliminate LBG
candidates with a brighter companion without a spectroscopic redshift due to fiber collisions.
The only difference between these earlier works and this one is that the LBG isolation criteria were
previously imposed using a WMAP7 cosmology \citep{2011ApJS..192...18K}, but here we use the cosmology of
\cite{2014A&A...571A..16P}.  Imposition of these selection criteria left 279~343 LBGs, of which we
use the} 249~818 that
lie within the area covered by the SDSS source catalog described in Sec.~\ref{subsec:source}.

We divide the galaxy sample into red and blue LBGs based on the $g-r$ color \revsw{K}-corrected to $z=0.1$,
using
\beq
^{0.1}g - ^{0.1}r = 0.8
\eeq
for our dividing line.  
\refresponse{As discussed in \cite{2015arXiv150906758Z}, the adoption of a stellar mass-dependent
  color cut to better isolate the red sequence modifies the fractions of galaxies in the red and
  blue subsamples by at most 3 per cent for the highest and lowest stellar mass subsamples, so the
  choice of a constant dividing line has little impact on our results.  Finally, we require redshift
  $z\ge0.03$, excluding the very small fraction of the sample at low redshift where the high galaxy
  flux at fixed luminosity induces more significant systematic uncertainty in sky subtraction and measurements of galaxy properties.}

\begin{figure*}
\begin{center}
\includegraphics[width=\textwidth,angle=0]{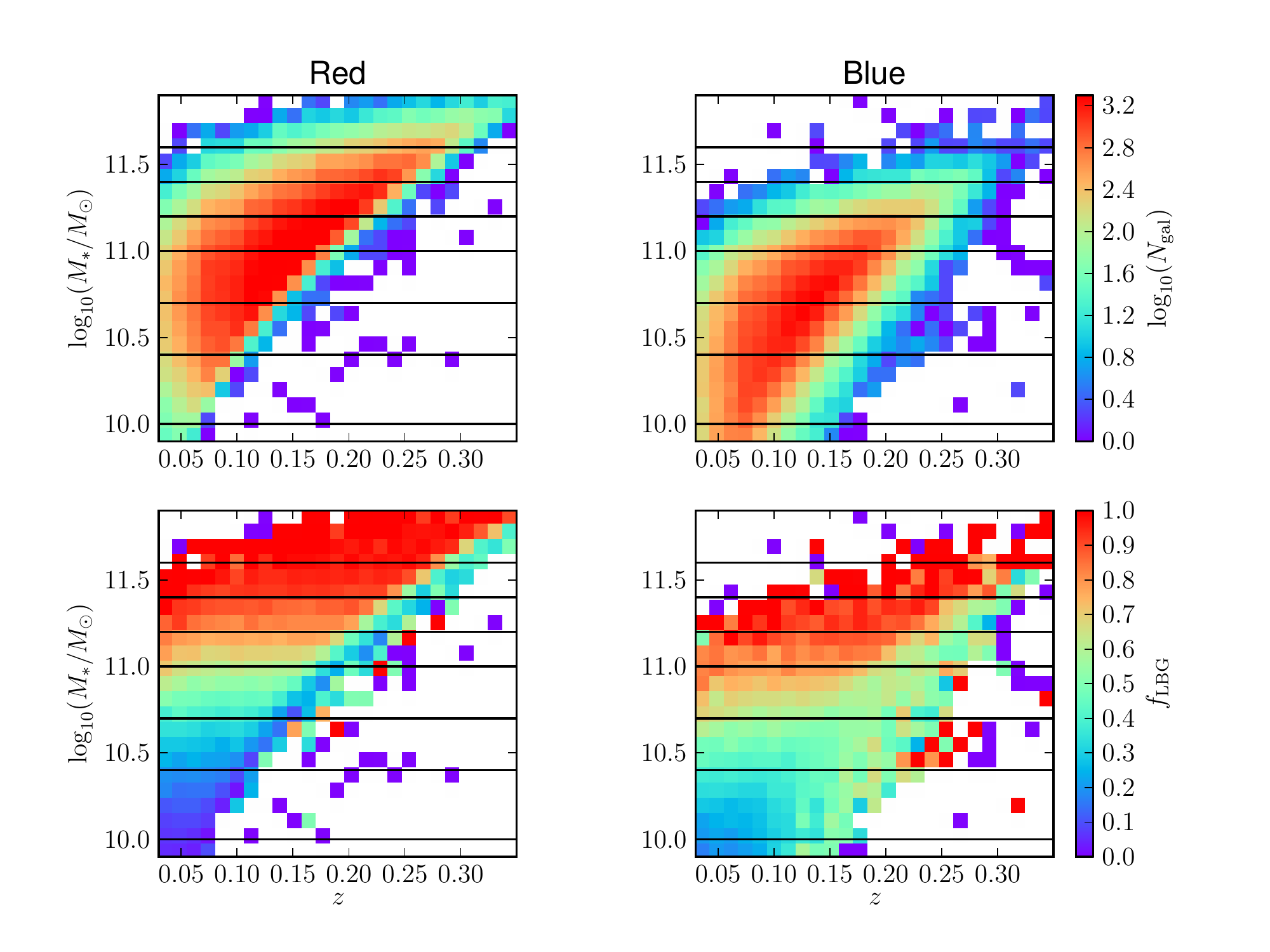}
\caption{\label{fig:sample}Top and bottom rows show \revy{the log of the number of LBGs} and the fraction of Main
  sample galaxies that are selected as LBGs, respectively, as a function of redshift and stellar
  mass.  Results for red and blue galaxies are shown separately in the left and right columns.  The
  color scales are defined self-consistently across the two columns.  Horizontal
  lines indicate the edges of the bins in stellar mass used for our analysis.}
\end{center}
\end{figure*}

Since the interpretation of stacked lensing signals can be complicated in cases where the mass
distribution is extremely broad, we define relatively narrow bins in stellar mass. \revsw{The halo
  mass distribution will still be broad due to intrinsic scatter between stellar and halo mass, but
  it improves the situation compared to that with very broad stellar mass bins.} However, 
relatively few high stellar mass LBGs are classified as blue, so we also present results for one very broad bin of blue galaxies above a stellar mass threshold of $10^{11}~M_\odot$.  The properties of the LBG
samples in each stellar mass and color bin are summarized in Table~\ref{tab:sample}.  In addition,
the number of galaxies and the LBG fraction as a function of redshift $z$ and stellar mass $M_*$ \revsw{are}
shown in Fig.~\ref{fig:sample}.  The table and figure clearly indicate the scarcity of blue LBGs 
at high $M_*$.  As shown, the fraction of Main sample galaxies that are LBGs is highest at
high stellar mass.

\revall{Fig.~\ref{fig:sample} illustrates that the sample is quite far from volume-limited.  We
  investigated the importance of the non-volume-limited nature of the sample by calculating LBG
  lensing signals for $z<0.08$, resulting in substantial reduction in sample size for all stellar mass
  bins except the first two (which are still not quite volume-limited after making this cut).
  However, while the errorbars increased significantly, there is no 
  strong modification in the lensing signal, just a slight shallowing below
  100~$h^{-1}$kpc. \refresponse{For a direct comparison of the lensing signals between the full LBG sample
    and the $z<0.08$ subsample, see Appendix~\ref{app:vollim}.} As
  discussed in Sec.~\ref{subsec:systematics}, these
  differences do not systematically change mass estimates in a way that suggests that a lack of
  volume-limited sample is driving our results, so for the rest of this paper we proceed with
  the flux-limited samples described above.} 

Since many previous papers that have explored the relationship between stellar and halo mass used
the MPA/JHU stellar mass 
estimates\footnote{\url{http://www.mpa-garching.mpg.de/SDSS/}}
(an updated version of the \citealt{2003MNRAS.341...33K} catalog) 
instead of those from the VAGC, Table~\ref{tab:sample} includes the effective mean stellar mass for
each sample using the MPA/JHU masses (with a color- and stellar mass-dependent conversion applied based
on comparison between the catalogs).  The differences are typically 0.1~dex for red galaxies, and
slightly less for blue galaxies, with the correction being larger at low stellar mass.  
While we adopt the VAGC stellar masses as our canonical ones used for
the analysis, to facilitate the comparison with previous work we will plot the results in
Sec.~\ref{subsec:comparison} as a function of the MPA/JHU masses, denoted $M_*^{(\text{MPA})}$.

\begin{table*}
  \begin{center}
    \caption{\label{tab:sample}Summary of the properties of the galaxy samples included in each
      color and stellar mass bin in this analysis.  The quantities that are tabulated are the
      minimum and maximum stellar mass, the number of galaxies $N_\text{gal}$, the effective
      weighted stellar mass of the galaxies taking into account their effective weight in the
      lensing measurement for our canonical (VAGC) stellar masses and the MPA/JHU stellar
      masses, the effective weighted redshift $z_\text{eff}$ of the galaxies in the bin
      including the lensing weight, and the fraction of Main sample galaxies $f_\text{LBG}$ in this
      bin that are selected as LBGs.}
    \begin{tabular}{ccccccc}
\hline\hline
$\log_{10}\left(\frac{M_{*,\mathrm{min}}}{M_\odot}\right)$ & $\log_{10}\left(\frac{M_{*,\mathrm{max}}}{M_\odot}\right)$ & $N_{\mathrm{gal}}$ &$\log_{10}\left(\frac{M_{*,\mathrm{eff}}}{M_\odot}\right)$ & $\log_{10}\left(\frac{M_{*,\mathrm{eff}}^{\text{MPA}}}{M_\odot}\right)$ & $z_{\mathrm{eff}}$ & $f_{\mathrm{LBG}}$ \\
\hline
\multicolumn{7}{c}{Red} \\
10.0 & 10.4 & \refresponse{4244} & 10.28 & 10.39 & 0.064 & 0.13 \\
10.4 & 10.7 & \refresponse{17542} & 10.58 & 10.70 & 0.081 & 0.27 \\
10.7 & 11.0 & \refresponse{44724} & 10.86 & 10.97 & 0.105 & 0.47 \\
11.0 & 11.2 & \refresponse{37987} & 11.10 & 11.20 & 0.131 & 0.66 \\
11.2 & 11.4 & \refresponse{28008} & 11.29 & 11.38 & 0.159 & 0.79 \\
11.4 & 11.6 & \refresponse{12599} & 11.48 & 11.56 & 0.191 & 0.88 \\
11.6 & 15.0 & \refresponse{3195} & 11.68 & 11.75 & 0.230 & 0.91 \\
\multicolumn{7}{c}{Blue} \\
10.0 & 10.4 & \refresponse{20690} & 10.24 & 10.29 & 0.079 & 0.32 \\
10.4 & 10.7 & \refresponse{30842} & 10.56 & 10.63 & 0.100 & 0.48 \\
10.7 & 11.0 & \refresponse{33621} & 10.85 & 10.94 & 0.124 & 0.65 \\
11.0 & 11.2 & \refresponse{11040} & 11.10 & 11.18 & 0.155 & 0.79 \\
11.2 & 11.4 & \refresponse{2626} & 11.28 & 11.35 & 0.183 & 0.87 \\
11.4 & 11.6 & \refresponse{325} & 11.47 & 11.54 & 0.220 & 0.90 \\
11.6 & 15.0 & \refresponse{96} & 11.68 & 11.69 & 0.246 & 0.96 \\
\hline
\end{tabular}

  \end{center}
\end{table*}

\subsection{Source sample}\label{subsec:source}

To measure the galaxy-galaxy lensing signal, we use a catalog \citep{2012MNRAS.425.2610R} of 1.2
background galaxies per arcmin$^2$ with galaxy shape measurements (corrected for the effects of the
point-spread function) estimated using the re-Gaussianization method \citep{2003MNRAS.343..459H} and
photometric redshifts from Zurich Extragalactic Bayesian Redshift Analyzer
\citep[ZEBRA,][]{2006MNRAS.372..565F}.  The catalog is characterized in detail in several papers
\citep[see ][]{2012MNRAS.425.2610R,2012MNRAS.420.1518M,2012MNRAS.420.3240N,2013MNRAS.432.1544M},
with \revsw{well-understood systematic uncertainties}.

\section{Mock catalogs}\label{sec:mocks}

We use one particular Munich semi-analytic model (SAM), a simulation of the development of the
galaxy distribution self-consistently including evolution and the gravitational environment, 
to create mock catalogs that are used to 
calibrate the mass biases and satellite contamination. The formation and evolution of galaxies are
modelled following the physics in the \cite{2011MNRAS.413..101G} model. \revall{Central galaxies are
  located at the center of the friends-of-friends (FOF) halos, while satellite galaxies are
  in subhalos that are not at the center of FOF halos and may include orphan galaxies for which the dark matter subhalo
  was disrupted before the present time.}   The stellar mass and
photometric properties of model galaxies are determined using stellar population synthesis models
from \cite{2003MNRAS.344.1000B} with the \cite{2003PASP..115..763C} initial mass function. The
uncertain star formation and feedback efficiencies were tuned to produce close fits to the stellar
mass function of low-redshift galaxies from SDSS. We further used an abundance matching-based method
to correct \revsw{the stellar masses by small amounts in order to bring the model stellar mass
  function into exact agreement with that in the SDSS (see \revall{\citealt{wenting-tmp}} where this
  particular model is denoted G11-P').} 
 The model also reproduces the observed luminosity and auto-correlation functions
of galaxies \refresponse{\citep{wenting-tmp}}.  

The original \cite{2011MNRAS.413..101G} model is based on the Millennium Simulation
\citep[][MS]{2005Natur.435..629S}, with WMAP1 $\Lambda$CDM cosmological parameters
\citep{2003ApJS..148..175S}. In our analysis, we scale the original MS to the first-year Planck
cosmology \citep[$H_0=67.3$, $\Omega_m=0.315$ and $\Omega_\Lambda=0.685$]{2014A&A...571A..16P}, by
adopting the scaling algorithm developed by \cite{2010MNRAS.405..143A} and improved by
\cite{2015MNRAS.448..364A}. Thus, in the end, the model has the same physics and parameters as
\cite{2011MNRAS.413..101G} but is based on \revsw{the} Planck cosmology. \refresponse{As shown in
  \cite{2010MNRAS.405..143A}, the rescaling algorithm can (in most respects) accurately reconstruct the matter
  field and properties of dark matter halos as they would have been simulated with another
  cosmological model; however, the concentrations of dark matter halos are less
  well-reproduced.  We show the sensitivity of our results to the assumptions in the halo
  concentration vs.\ mass relation in Sec.~\ref{subsec:systematics}.}

\subsection{\refresponse{Mock LBG samples}}

We construct mock LBG catalogs using exactly the same selection criteria as for SDSS LBGs. We first
project the simulation box along the $z$-axis (representing the line-of-sight direction). Each
galaxy can be assigned a redshift based on its $z$ coordinate and velocity. Selections can then be
made based on the projected separation and redshift difference in the same way as in the
observational data. The direct projection of the simulation box maximises the statistical signal,
but fails to incorporate observational effects such as the flux limit of the real survey, the
\revsw{K}-corrections to obtain rest-frame magnitudes, and the incompleteness of close pairs caused by fibre
collisions and the complex geometry of SDSS. \cite{2012MNRAS.424.2574W} and
\cite{2014MNRAS.442.1363W} have compared satellite properties based on such direct projections of
the simulation box to a full light-cone mock catalogue, and they found that the direct projection gives
unbiased results. For our later analysis we will use both LBGs in the $z=0$ snapshot and those in
earlier snapshots to account properly for the evolution of \revsw{galaxies and their halos.}
\revfix{However, we have confirmed that the halo mass distributions as a function of stellar mass
  and color do not differ very strongly between the $z=0$ snapshot and the multi-snapshot catalogs.}

To further validate our use of these mock catalogs for this study, in Sec.~\ref{subsec:lensing}, we
will compare the lensing profiles of the simulated LBG samples in stellar mass and color subsamples
with those of real LBGs. \revsw{In \revall{\cite{wenting-tmp}}, the predicted dark
matter profiles from eight different semi-analytic models with varying cosmology and model physics
are compared with SDSS lensing profiles, and this particular model  is
the one} that gives the best match to SDSS. However, that comparison uses only stellar mass binning
of the sample, 
so our test here using color divisions may be considered a more stringent test.

\subsection{Completeness and purity of central galaxy samples}\label{subsec:comppur}

\subsubsection{Completeness}\label{sec:completeness}

First, we use the $z=0$ snapshot of the mock catalog to assess the completeness of the sample of
central galaxies selected using the LBG cut.  The completeness is quantified as the fraction of central
galaxies of a given stellar mass and color that pass the LBG selection criteria.  We also check whether
the LBG selection is independent of the host halo mass, i.e., is the typical halo mass of central
LBGs the same as the typical halo mass of central galaxies overall?  
Incompleteness that causes a bias in the halo mass distribution is more problematic for our
purposes.  For
``typical'' masses, we consider a straight average, a median, and $\langle M_{200m}^{2/3}\rangle^{3/2}$; since the
mass from an NFW fit scales like $\Delta\Sigma^{3/2}$ on small scales \citep[e.g.,][]{2010MNRAS.405.2078M},
averaging $\Delta\Sigma$ corresponds to averaging $M_{200m}^{2/3}$. 
However, since most methods measure $\langle M_{200m}\rangle$, we will use that as our canonical
statistic for comparison of masses throughout this work, and apply corrections to best-fitting
masses to get an average halo mass.

\begin{figure}
\begin{center}
\includegraphics[width=\columnwidth,angle=0]{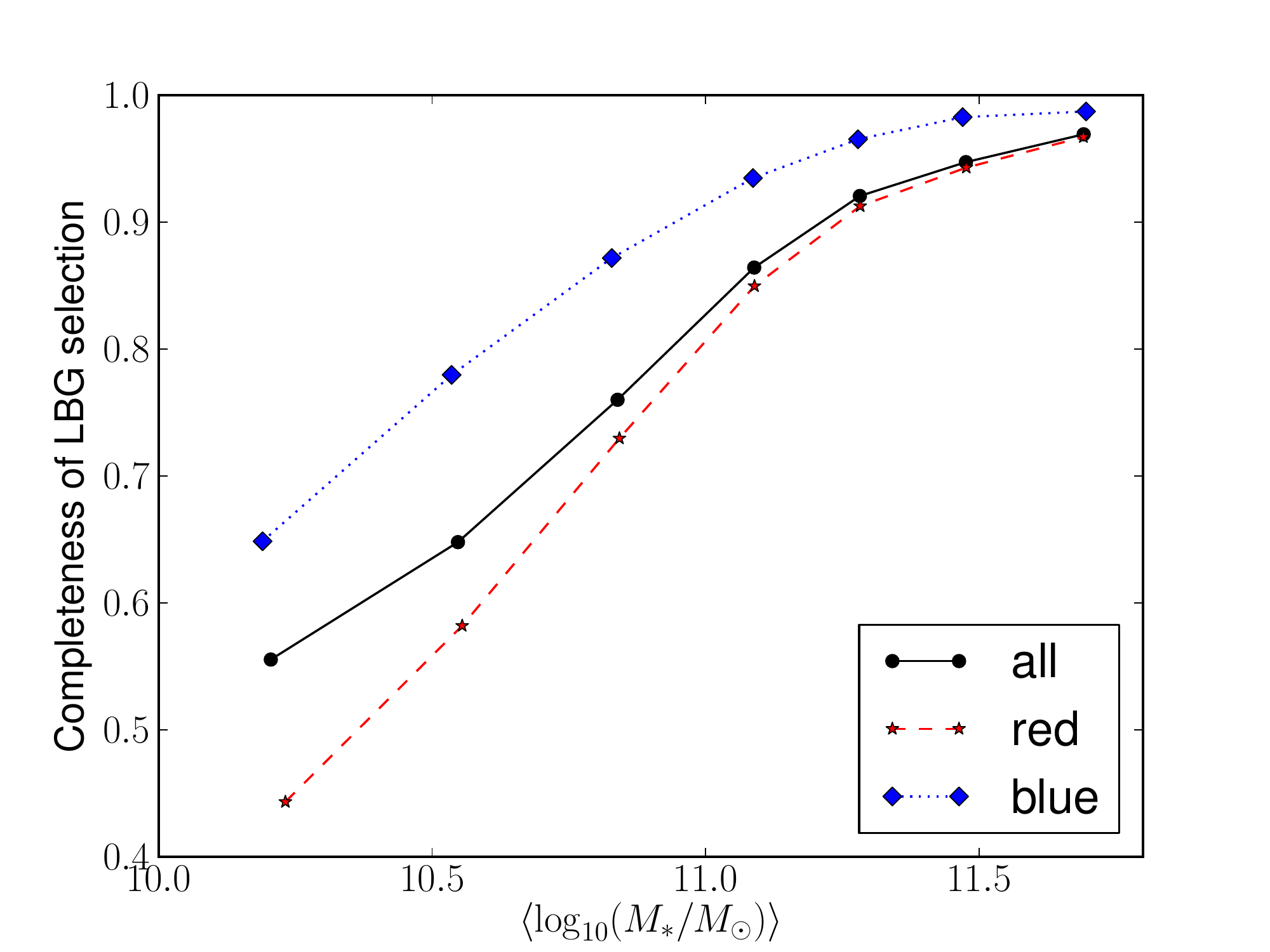}
\caption{\label{fig:compl}The completeness of the central galaxy selection as carried out using the
  LBG criterion in stellar mass bins, for red, blue, and all LBGs shown as red, 
blue, and black points, respectively.}
\end{center}
\end{figure}

\begin{figure}
\begin{center}
\includegraphics[width=\columnwidth,angle=0]{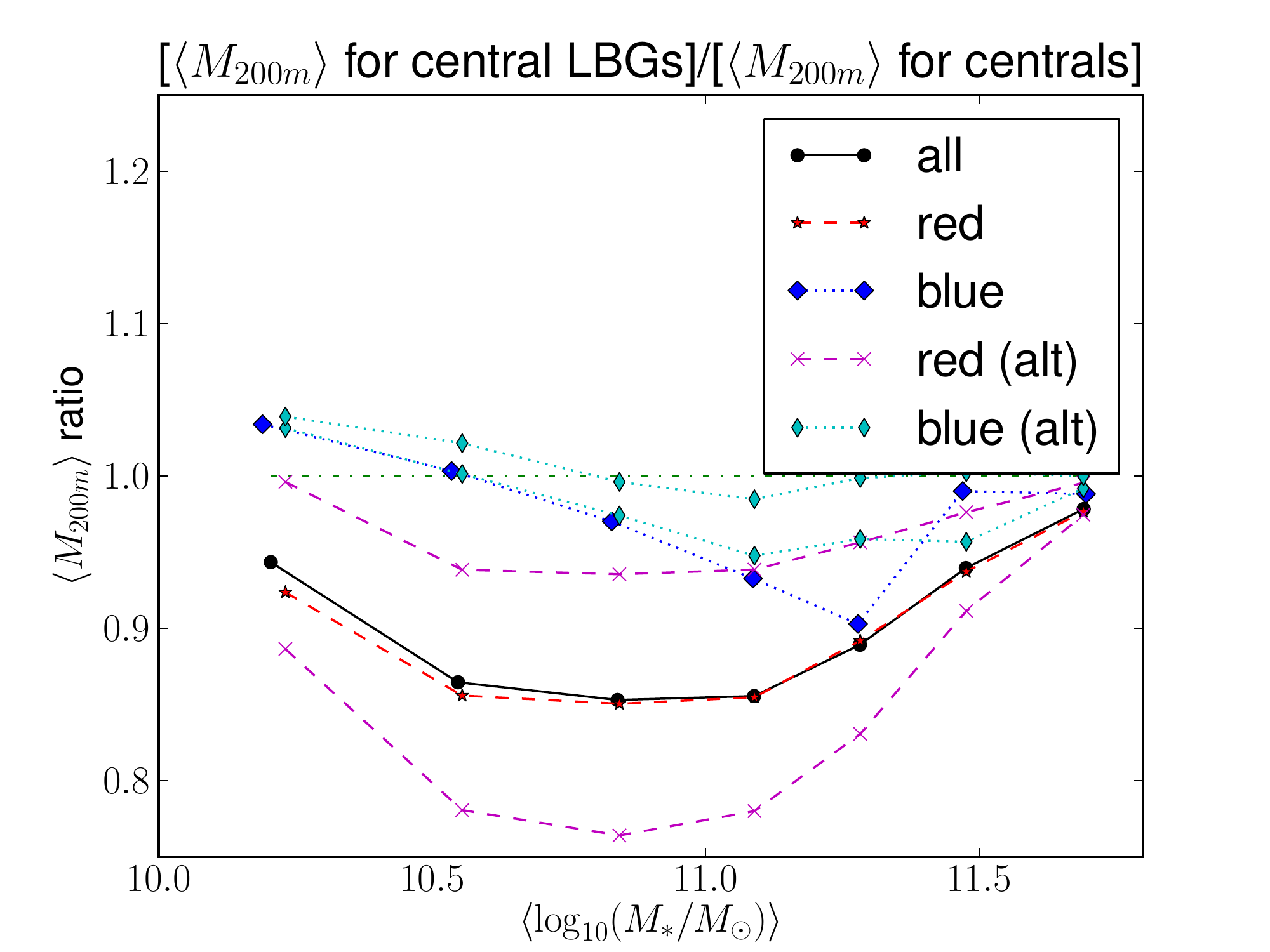}
\caption{\label{fig:mass_compl}The ratio of the mean halo mass of central LBGs to the mean halo mass
  of all centrals in stellar mass bins, for red, blue, and all LBGs shown as red, blue, and black
  points, respectively.  \refresponse{The additional magenta and cyan points labeled as ``alt'' in
    the legend were calculated using alternate mock catalogs that will be discussed in more detail
    in Sec.~\ref{subsec:systematics}, as a way of quantifying systematic uncertainty in these
    results.}}
\end{center}
\end{figure}

The completeness is shown in Fig.~\ref{fig:compl}, and the ratio of typical central LBG halo mass to
typical central galaxy halo mass is shown in Fig.~\ref{fig:mass_compl}. As Fig.~\ref{fig:compl}
illustrates, the completeness of the central galaxy sample selected using the LBG criterion is quite
high ($>90$ per cent) for stellar masses above $10^{11}~M_\odot$, but drops below that.  There are
two reasons for this incompleteness.  First, as discussed in \cite{2013A&A...557A..52P}, many of the
``missing'' central galaxies were excluded because they are not the brightest object in their own
halo (this explanation is also qualitatively consistent with the results of
\citealt{2011MNRAS.410..417S} and \citealt{2015MNRAS.452..998H}).
Second, the typical halo masses of the central galaxies with stellar mass below $10^{11}~M_\odot$ is
such that the $1$~Mpc cylinder used in the LBG selection exceeds the virial radius.  Hence low
stellar mass centrals can be excluded because of a brighter galaxy within a distinct neighboring
halo.  This effect is more important for red than for blue LBGs, presumably since red galaxies tend
to occupy more overdense regions.

However, for this study we are more concerned with the results in Fig.~\ref{fig:mass_compl}, which
indicate whether the mean halo mass of central LBGs differs significantly from the mean halo
mass of all centrals.  The former is what we measure, but the latter is what we wish to infer.
As shown, for blue galaxies the ratio of these mean halo masses always exceeds 0.9.
For red galaxies, which suffer from more incompleteness, the ratio can be as low as 0.85, meaning
that red central LBGs have masses that are biased low by about 15 per cent relative to those of all
red centrals at the same stellar mass.  The curves for ``red'' and ``all'' have similar trends that
arise due to their having broader halo mass distributions than the ``blue'' sample, so \revy{even} a slight
mass-dependence in the LBG selection can more easily modify the mean halo mass. \revall{Moreover,
  the ``red'' and ``all'' curves are similar because the contribution of red LBGs to the ``all''
  curve is boosted by the overall higher halo masses for red LBGs.} If the ratios in
Fig.~\ref{fig:mass_compl} had been very far from 1 (e.g., 0.5) then this
would call our entire method into question; but we will correct for this bias using the curves in
Fig.~\ref{fig:mass_compl}, relying on the realism of the mock catalogs to provide valid
corrections. 

\revy{We emphasize that Fig.~\ref{fig:mass_compl} rules out the idea that we might observe different halo
masses for the red vs.\ blue LBG samples solely due to selection effects (e.g., that the red LBG
sample might be heavily biased towards high-mass halos due to the selection procedure preferentially
excluding low-mass red centrals that are in overdense regions).} \revall{While the corrections
derived from Fig.~\ref{fig:mass_compl} are in principle dependent on the details of the mock
catalogs, we confirm in Sec.~\ref{subsec:systematics} that even significant modifications in the
mock catalogs do not lead to major changes in these corrections\refresponse{; alternative results for two
different mock catalogs discussed in that section are shown on Fig.~\ref{fig:mass_compl}.}}

Our finding that there is a small bias in the mean halo mass due to LBG selection is
consistent with the findings of \cite{2009MNRAS.392..801M}.  They compare the results of an
iterative, luminosity-dependent isolation criterion with a fixed aperture criterion (which is
stricter than the one used here, as it requires a factor of two difference in luminosity between the
LBG and anything else in its cylinder, instead of that it be strictly the brightest).  As shown
in their figure A1, the fixed isolation criterion can give a stellar mass-dependent suppression of the
mean halo mass of selected galaxies compared to that of all central galaxies at that stellar mass.
While a quantitative comparison is not possible due to the different cylinder parameters and
brightness requirements, our results are qualitatively in agreement with theirs.

\subsubsection{Purity}\label{sec:purity}

\begin{figure}
\begin{center}
\includegraphics[width=\columnwidth,angle=0]{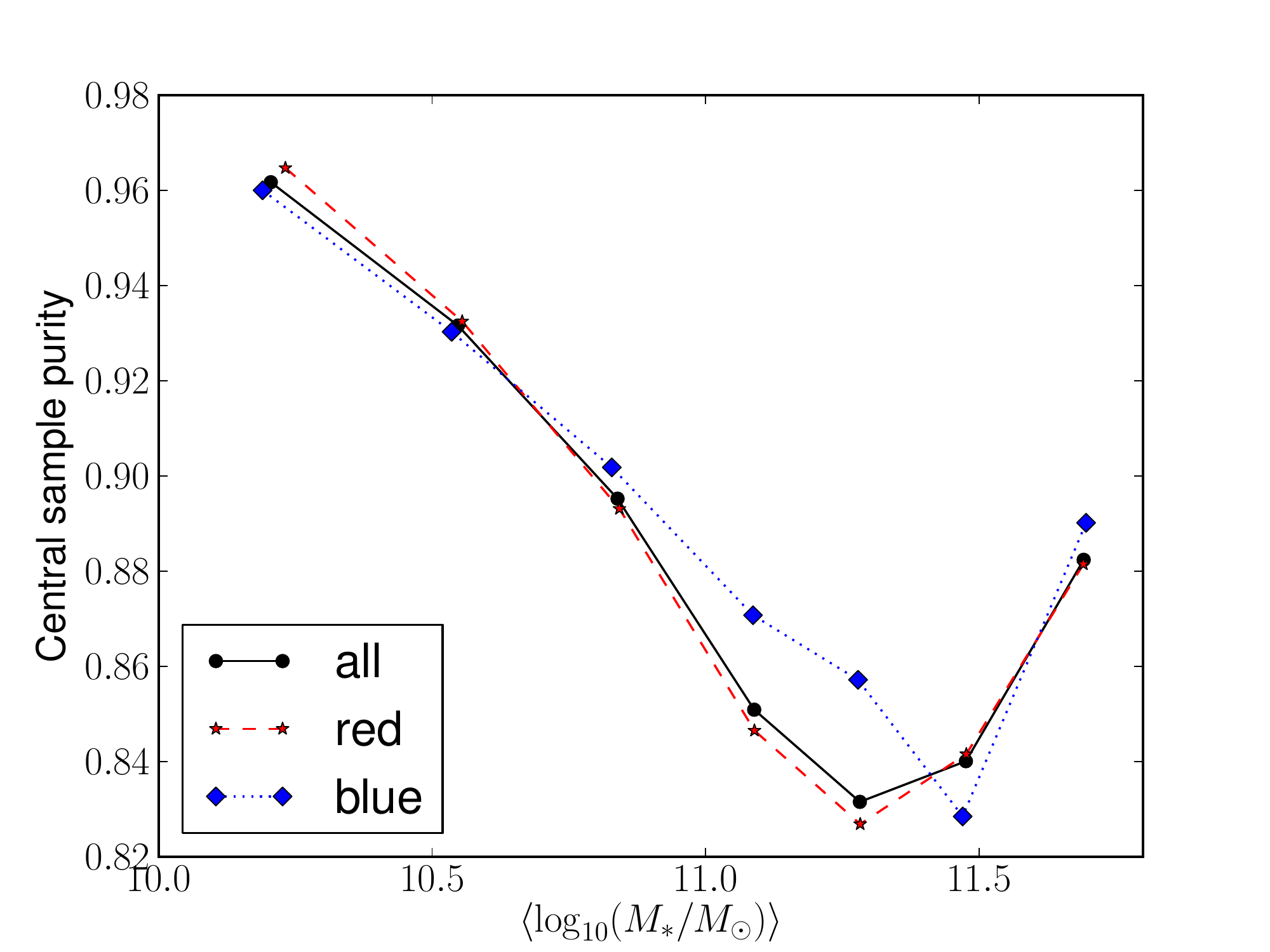}
\caption{\label{fig:purity}The purity of the central galaxy sample obtained using the 
  LBG criterion in stellar mass bins, for red, blue, and all LBGs shown as red, 
blue, and black points, respectively.}
\end{center}
\end{figure}

In addition, it is important to understand the purity of the LBG sample as a set of central
galaxies.  Contamination from satellites can modify the shape and increase the amplitude of the weak
lensing signal on scales above $\sim 300h^{-1}$kpc \citep[e.g.,][]{2005MNRAS.362.1451M}, affecting
mass estimates. This effect is particularly important if the contaminating satellites reside in very high-mass
halos.  Thus, Fig.~\ref{fig:purity} shows the purity of the LBG sample in the $z=0$
snapshot of the simulation, defined as the ratio of number of central LBGs to number of LBGs as a
function of stellar mass and color.  As shown, at stellar masses below $10^{11}~M_\odot$, the purity
of the central sample determined using the LBG selection \refresponse{exceeds $\sim 90$} per cent.  For higher stellar
mass, it can go as low as 82 per cent.  This is most likely related to the fact that (as mentioned in 
Sec.~\ref{sec:completeness}) the brightest galaxy in a halo is not always a central galaxy.

However, the  magnitude of the impurity alone cannot be used to quantify 
how the contaminating satellites bias the mass estimates from fits to
$\Delta\Sigma$.  The bias in the masses depends on the
details of the subhalos and larger host halos in which the contaminating satellites reside, the
distribution of their separations from the center of the host halo, and other effects.  For example,
in the case of contamination due to the brightest galaxy not being the central, it may often be the
case that the mass of the host halo is not very different from the mass one would expect if it had
been a central, so the contamination in the lensing signal and mass estimates will not be very
large.  In the next
section, we will quantify the implications of satellite contamination on  mass estimates 
directly, by fitting the lensing signal in the mock LBG samples (including the contaminating
satellites) to estimate masses, and directly determining the bias.


\section{Method}\label{sec:method}

\subsection{Measurements}

Galaxy-galaxy weak lensing, the coherent tangential shape distortion (or shear) of background
galaxies due to the matter in foreground lens galaxies, probes the connection between the lens
galaxies and matter via their cross-correlation function $\xi_{gm}$.  This cross-correlation can be
related to the projected surface density\footnote{Here we neglect the very broad radial window
  function\refresponse{, given that it makes sub-$0.2$\% differences in the signals on the scales used in this
  work, and neglecting it enables faster computations of theoretical signals.}} via
\begin{equation}\label{E:sigmar}
\Sigma(r_p) = \overline{\rho} \int \left[1+\xi_{gm}\left(\sqrt{r_p^2 + \pi^2}\right)\right] \mathrm{d}\pi,
\end{equation}
where \revsw{$\overline{\rho}$ is the mean matter density,} $r_p$ is the physical projected separation and $\pi$ is the line-of-sight distance from the
lens.  The surface density is then related to the observable quantity for lensing, the surface
density contrast:
\begin{equation}\label{E:ds}
\Delta\Sigma(r_p) = \gamma_t(r_p) \Sigma_\text{crit}= \overline{\Sigma}(<r_p) - \Sigma(r_p).
\end{equation}
\revsw{Here $\overline{\Sigma}(<r_p)$ is the average surface density within $r_p$.  
The observable $\Delta\Sigma$} is the product of two factors, a tangential shear $\gamma_t$ and a geometric factor
\begin{equation}\label{E:sigmacrit}
\Sigma_\text{crit} = \frac{c^2}{4\pi G} \frac{D_S}{D_L D_{LS}}
\end{equation}
where $D_L$ and $D_S$ are angular diameter distances to the lens and source, $D_{LS}$ is the angular
diameter distance between the lens and source, and physical (rather than comoving) coordinates are used for all quantities
throughout this paper.

The lensing measurement begins with identification of lens-source pairs, with source photometric
redshift larger than the lens spectroscopic redshift.  Inverse variance weights are assigned to each
lens-source pair, including both shape noise and measurement error terms in the variance:
\begin{equation}
w_{ls} = \frac{1}{\Sigma_{\rm crit}^{2}(\sigma_e^2 + \sigma_{\rm{SN}}^2)},
\label{eq:wls}
\end{equation}
where $\sigma_e^2$ is the shape measurement error due to pixel noise, and $\sigma_{\rm{SN}}^2$ is
the root-mean-square intrinsic ellipticity (both quantities are per component, rather than total;
the latter is fixed to $0.365$ following \citealt{2012MNRAS.425.2610R}).  Use of photometric
redshifts which have nonzero bias and significant scatter results in a bias in $\Delta\Sigma$ that
can be easily corrected using the method from \cite{2012MNRAS.420.3240N}.  This bias is a function
of lens redshift, and is calculated including all weight factors for each lens sample taking into
account its redshift distribution.

$\Delta\Sigma$ in bins in $r_p$ can be computed via a summation over lens-source pairs ``$ls$'' and
random lens-source pairs ``$rs$'':
\begin{equation}
\Delta\Sigma(r_p) = \frac{\sum_{ls} w_{ls} e_t^{(ls)} 
\Sigma_{{\rm crit}}(z_l,z_s)}{2 {\cal
    R}\sum_{rs} w_{rs}},
\end{equation}
where $e_t$ is the tangential ellipticity component of source galaxy with respect to the lens
position, and $\mathcal{R}$ is the shear responsivity \citep{2002AJ....123..583B} that converts from
the ensemble average distortion to shear.  The division by $\sum w_{rs}$ accounts for the fact that
some of our ``sources'' are physically associated with the lens, and therefore not lensed by it
\citep[see, e.g.,][]{2005MNRAS.361.1287M}.  Finally, we subtract off a similar signal measured
around random points with the same area coverage and redshift distribution as the lenses, to
subtract off any coherent systematic shear contributions \citep{2005MNRAS.361.1287M}; this signal is
statistically consistent with zero for all scales used in this work.

To calculate the error bars on $\Delta\Sigma$, we use the bootstrap resampling
method, dividing the lens sample into 100 patches on the 
sky. Previous work \citep[e.g.,][]{2005MNRAS.361.1287M,2013MNRAS.432.1544M} has shown that for the
scales used here, the errorbars are dominated by uncorrelated shape noise (whereas on larger scales,
correlated shape noise, systematics, and cosmic variance result in non-negligible correlations
between bins in $r_p$).  For this reason, we use only diagonal errors throughout this work.  The
maximum scale used for the fits in the lensing  
analysis is 1~\hmpc, which for a typical lens redshift is far below the typical size of each
jackknife resampled region.  Thus, the bootstrap method is a reasonable approach to getting the
covariance matrix for the projected mass profile, and has been shown to agree with several other
approaches in the shape noise-limited regime \citep{2005MNRAS.361.1287M}. 

\subsection{Mass estimation}\label{subsec:massest}

When estimating masses, we take advantage of the simplicity of the LBG sample as a nearly pure
sample of central galaxies.  All of the projected mass around these lens galaxies contributes to
  the galaxy-galaxy lensing signal.  This includes contributions from the host dark matter halo in
  which the lens galaxy resides (``1-halo term''), and from other dark matter halos (``two-halo
  term'') that are part of large-scale structure associated with the lens.  For central galaxies,
  the 1-halo term simply corresponds to the $\Delta\Sigma$ for the host dark matter halo.  For
  satellite galaxies, there are two contributions to the 1-halo term: on small scales, a
  contribution from the satellite subhalo, and on larger scales ($0.3$--$2h^{-1}$Mpc) a contribution
  from the host halo itself.  For a nearly pure sample of central galaxies, we can restrict
  ourselves to the 1-halo regime and take advantage of the fact that the halo profiles of centrals
  are approximately NFW \citep{1997ApJ...490..493N} profiles.  

For all but the highest mass bin, we use a minimum
  radius of $r_p=50$\hkpc\ (physical), whereas for the highest mass bin where there is
  evidence of large-scale light profiles interfering with the measurements of the nearby fainter
  galaxies used for lensing measurements, our minimum scale is $r_p=70$\hkpc.  \revall{The
    contribution of the galaxy stellar mass to the lensing signal is negligible compared to the dark
  matter halo contribution for $r_p$ above our minimum value.} In all cases, the
  maximum scale is 1~\hmpc, ranging from several times the virial radius to only a small amount
  outside the virial radius. \refresponse{We rely on the calibration factors derived from using the
    same fitting procedure on mock catalogs to correct for any small contamination of the signal on
    these scales by the two-halo term, which may be more important at lower mass (where we go
    farther beyond the virial radius).}

  Our fits have a fixed concentration-mass relation, corresponding to that from \cite{2015ApJ...799..108D}, and
  evaluated at the lensing-weighted mean redshift of each sample.  However, when presenting our results we will
  estimate the sensitivity of our results to assumptions about the concentration-mass relation. The
  masses quoted in this paper correspond to a spherical overdensity of $200$ with respect to the
  mean density ($M_{200m}$), using \cite{2014A&A...571A..16P} cosmological parameters ($\Omega_m=0.315$).

  As shown in \cite{2005MNRAS.362.1451M}, using fits to NFW profiles when the signal is actually the average of many
  NFW profiles spanning a range of masses and concentrations can give a best-fitting mass that is biased with respect
  to the true mean mass by a few to tens of per cent (depending on the width of the
  mass distribution).  \revsw{This point is quite relevant given that \cite{2013A&A...557A..52P} showed that the typical FWHM of the halo mass distribution at fixed
  stellar mass for LBGs is $\sim 1$ dex.}  In addition, NFW fitting can result in mass
  biases if the assumed concentration is incorrect, or if the fit includes scales outside the virial
  radius where the simple extrapolation of the NFW profile may be incorrect.  Our approach in this
  work is to use $N$-body and SAM-based mock catalogs to calibrate out such 
  biases as well as those due to satellite contamination in our LBG sample (see 
  Sec.~\ref{sec:mocks}).

Errorbars on the mass estimates are
calculated using the bootstrap.  Each bootstrap-resampled $\Delta\Sigma$ is fit using the
procedure described in this section, and the quoted results are the median of the mass
distribution from all bootstrap-resampled datasets, with errorbars indicating the 16 and 84 percentiles.  For all but the lowest mass bin
(for red and blue) 
and the two highest-mass blue samples,
where the signal-to-noise is lower, using these percentiles gives nearly identical errors to using
the standard deviation.  Also, the median of the bootstrap masses is consistent with the mass obtained by
directly fitting the average signal to within $\lesssim 0.1\sigma$.  It should be kept in mind that
we are quoting percentiles of the distribution of mean halo mass values from the bootstrap; these
are errors on the mean mass, not a measure of scatter in the halo mass distribution (which is much
broader than our quoted errors on the mean).

\revsw{  To summarize, we are inferring $\langle M_{200m}\rangle (M_*)$, the average halo
  mass in bins of stellar mass.  This is the most observationally accessible quantity in a lensing
  measurement.  Other works describe the relationship between stellar and
  halo mass in other ways.  For example, halo models commonly infer a stellar-to-halo mass relation
  (SHMR) parametrized as $M_*(M_{200m})$ which is a continuous function representing the median
  stellar mass of central galaxies occupying halos of fixed halo mass.  In principle, the quantity
  that we measure in this work for red and blue central
  galaxies depends on all of the following: the continuous SHMR, the size of the intrinsic and
  observational scatter about the median SHMR, the halo mass function, and any halo mass dependence
  in the fraction of red vs.\ blue galaxies.  It is important to bear in mind the distinction between the
  observationally-accessible $\langle M_{200m}\rangle (M_*)$ and the underlying SHMR when comparing
  our results with others that infer a parametrized $M_*(M_{200m})$.}

\subsection{Fidelity of mass estimation procedure}\label{subsec:massfidelity}

Next, we consider the fidelity of the mass estimates from our NFW fitting procedure.  The average of
a collection of NFW profiles does not look precisely like an NFW profile with the average mass, and simplifications in the assumed
mass-concentration relation along with deviations from an NFW profile extrapolated beyond the virial
radius can cause biases.  To quantify these effects, we applied
our mass fitting procedure to stacked $\Delta\Sigma$ profiles for {\em central} LBGs selected across
multiple redshift snapshots in the mock catalogs.

A second complication is that satellites that are accidentally included in the LBG sample, which are not
completely negligible (Sec.~\ref{sec:purity}), modify the $\Delta\Sigma$ profiles
on intermediate scales and bias the mass estimates.  To address this point, we applied our mass
fitting procedure to stacked $\Delta\Sigma$ profiles for {\em all} (not just central) LBGs selected across multiple
redshift snapshots \revfinal{while matching the joint redshift-stellar mass distribution of the real
  LBG
  sample, and including redshift-dependent weights consistent with the lensing analysis in the real
  data as in \cite{wenting-tmp}}.  This test also includes the issues covered by
the first test, but carrying out both tests separately allows us to assess the relative importance of satellite
contamination vs.\ other issues.

\begin{figure}
\begin{center}
\includegraphics[width=\columnwidth,angle=0]{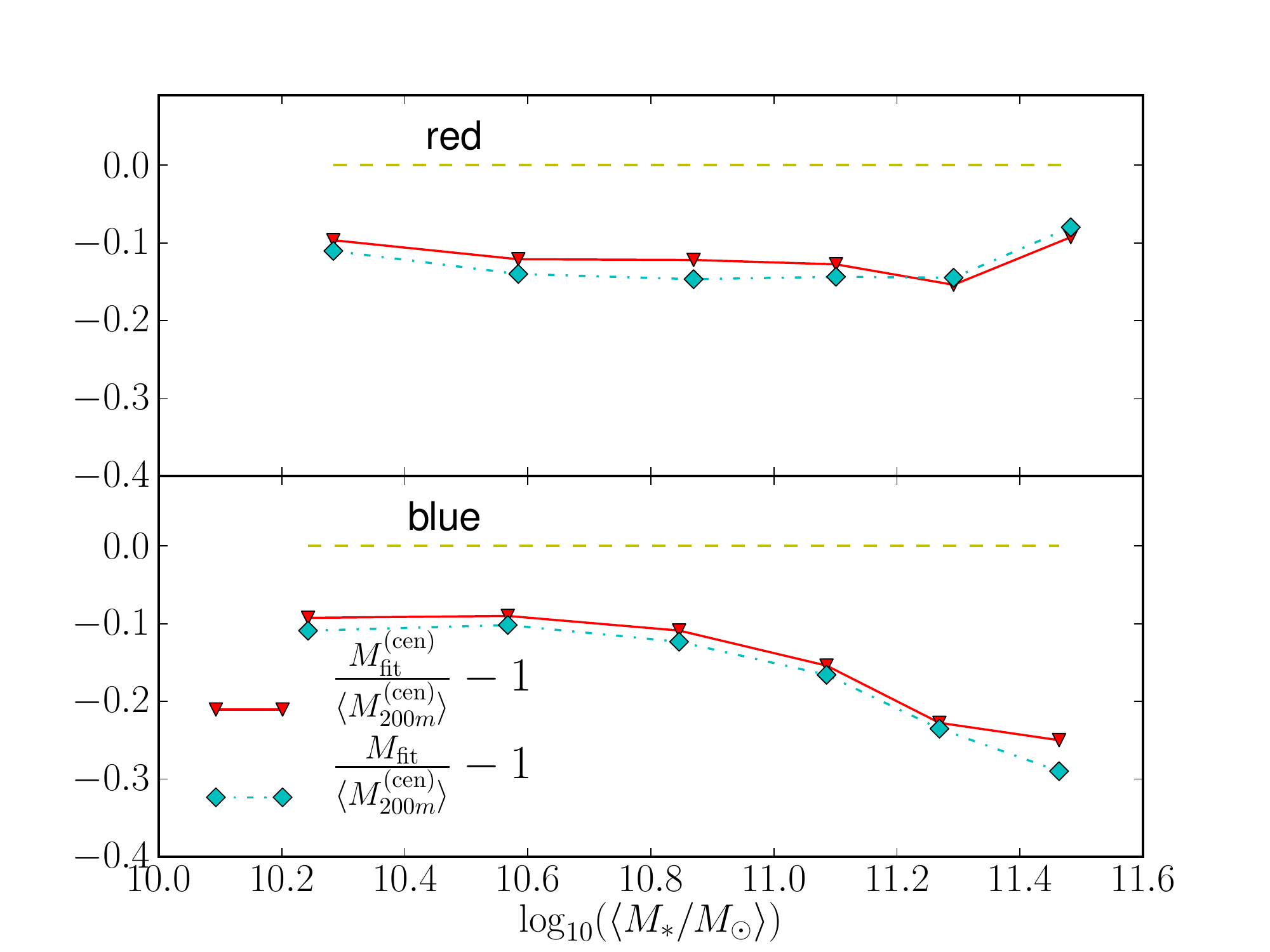}
\caption{\label{fig:massbias}The bias in the best-fitting NFW mass compared to the mean
  halo mass
  for central LBGs, for red (top), and blue (bottom) LBGs respectively.  The two lines
  that are shown indicate the bias in the estimated masses of central LBGs, which result from the
  fitting procedure, and in the estimated masses of all LBGs (including those that are not central,
  thus including the effects of satellite contamination).  The ideal unbiased case is shown as a
  horizontal dashed line.}
\end{center}
\end{figure}

\begin{figure}
\begin{center}
\includegraphics[width=\columnwidth,angle=0]{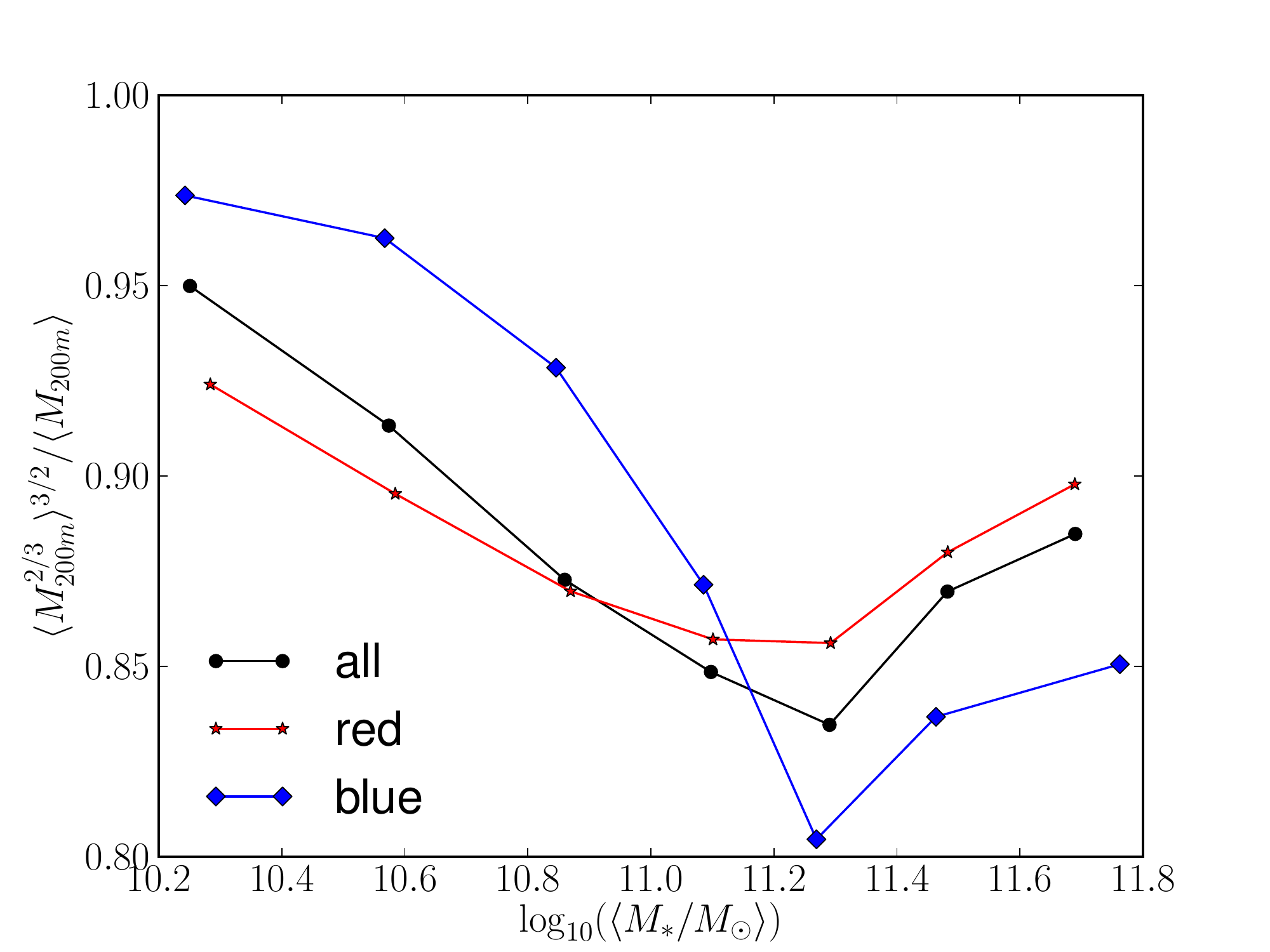}
\includegraphics[width=\columnwidth,angle=0]{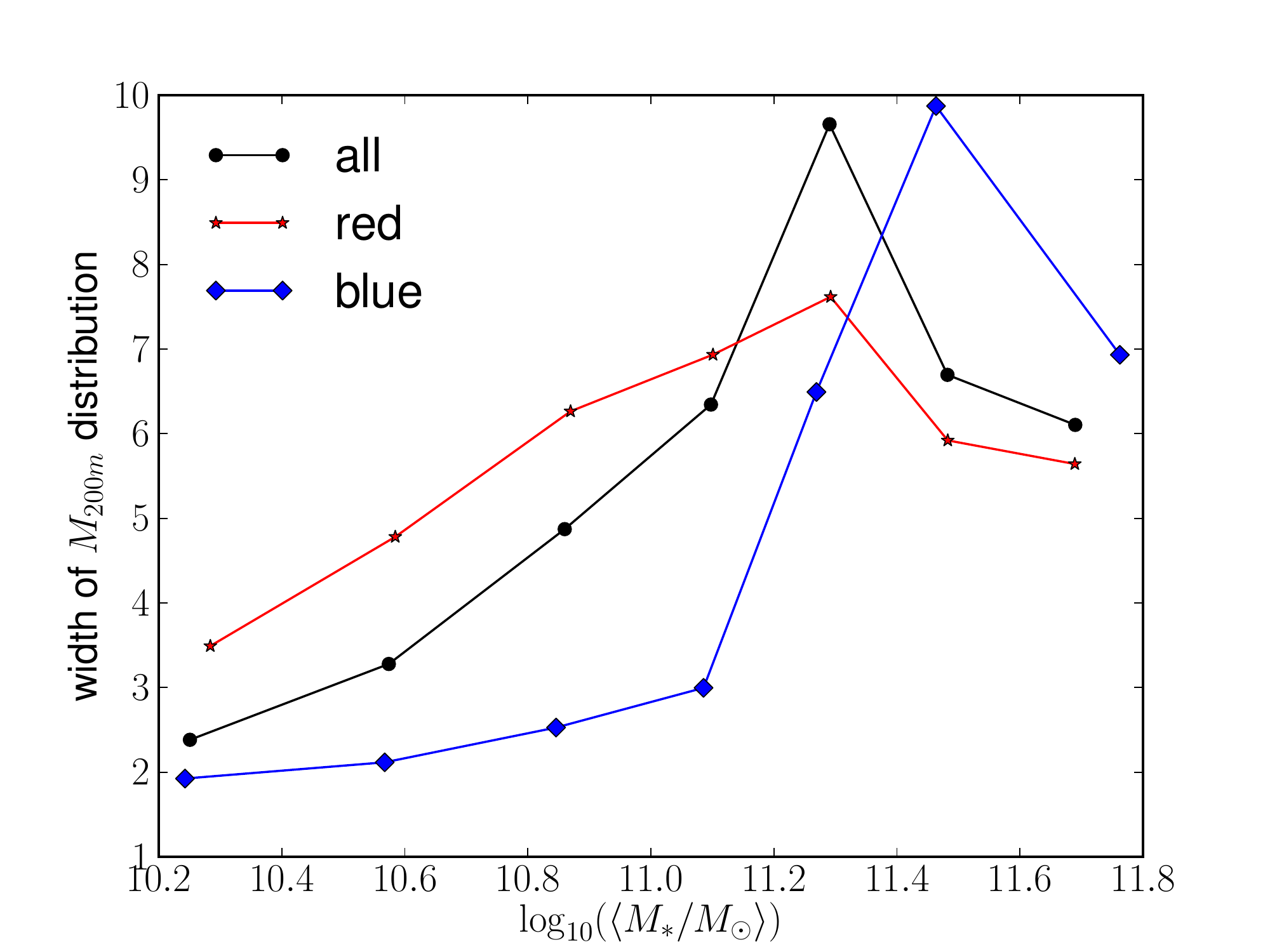}
\caption{\label{fig:hm_dist}{\em Top:} The ratio of the lensing signal-weighted average mass
($\langle M_{200m}^{2/3}\rangle^{3/2}$) to the mean mass $\langle M_{200m}\rangle$ for central LBGs in the mock
catalog.  {\em Bottom:} \revsw{The width of the halo mass distribution, quantified as the ratio of the halo masses corresponding to the 84th and 16th percentile
of the central LBG halo mass distribution.}}
\end{center}
\end{figure}

Results are
shown in Fig.~\ref{fig:massbias}, where we plot \refresponse{the bias of the best-fitting masses
  with respect to the true average halo mass for central galaxies in this stellar mass bin,
  defined as the ratio of the best-fitting mass to the average central galaxy halo 
mass, minus 1.  For the overall LBG sample, this is denoted $M_\text{fit}/\langle M_{200m}^\text{(cen)}\rangle-1$, while for central LBGs it is denoted
$M_\text{fit}^\text{(cen)}/\langle M_{200m}^\text{(cen)}\rangle-1$.}  As shown, the bias tends to be negative, i.e., an
underestimation of the mean mass.  This is also true if we use the lensing signal-weighted average mass
($\langle M_{200m}^{2/3}\rangle^{3/2}$), which makes the bias slightly less severe.  Our finding that the
best-fitting mass is below the mean mass is consistent with the results of
\cite{2005MNRAS.362.1451M}, who found that when assuming a single value of mass, the best-fitting
mass tended to be below the mean mass and above the median mass.  In that work, 
the bias was worse when the mass distribution was broader.  Fig.~\ref{fig:hm_dist} shows that this
is the case here too\revfix{, though the trend with breadth of the halo mass distribution is weak}; the bins in Fig.~\ref{fig:massbias} that
have a larger bias \revfix{for blue LBGs at high stellar mass} have a
broader halo mass distribution, also resulting in a bigger difference between $\langle
M_{200m}\rangle$ and $\langle M_{200m}^{2/3}\rangle^{3/2}$.

The difference between the results for the two tests shown in Fig.~\ref{fig:massbias} is very small, indicating that satellite
contamination is not significantly biasing the mass fits, perhaps for the reasons suggested in Sec.~\ref{sec:purity}.  Moreover, the biases in the masses for
red and blue galaxies are fairly similar, so they should not compromise our ability to
distinguish between \revsw{the halos of} red and blue galaxies.

\section{Results}\label{sec:results}

\subsection{Lensing signals}\label{subsec:lensing}

\begin{figure*}
\begin{center}
\includegraphics[width=6.4in,angle=0]{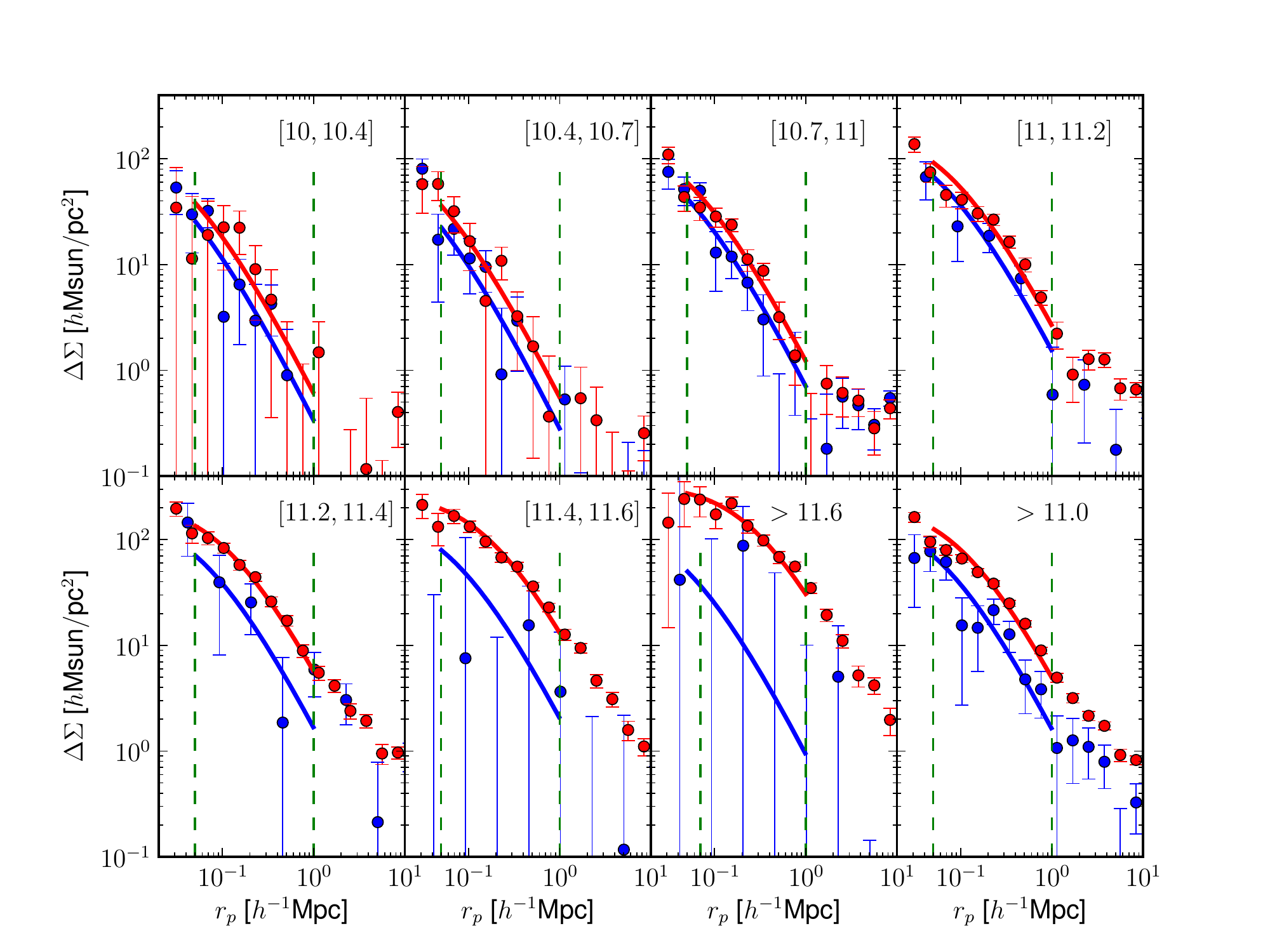}
\caption{\label{fig:signals}Lensing signals for LBGs in stellar mass bins, for red and blue LBGs shown as red
and blue points with errors, respectively.  \revy{The thick lines show the best-fitting NFW profiles, while the
vertical dashed \refresponse{lines show the lower and upper limits} in $r_p$ used for the fits.  For viewing purposes, the
data for blue LBGs in the four narrow bins
with $M_*>10^{11}~M_\odot$ have been rebinned into bins that are twice as
broad as those shown for red LBGs, due to the lower signal-to-noise ratio.}}
\end{center}
\end{figure*}

\begin{figure*}
\begin{center}
\includegraphics[width=6.4in,angle=0]{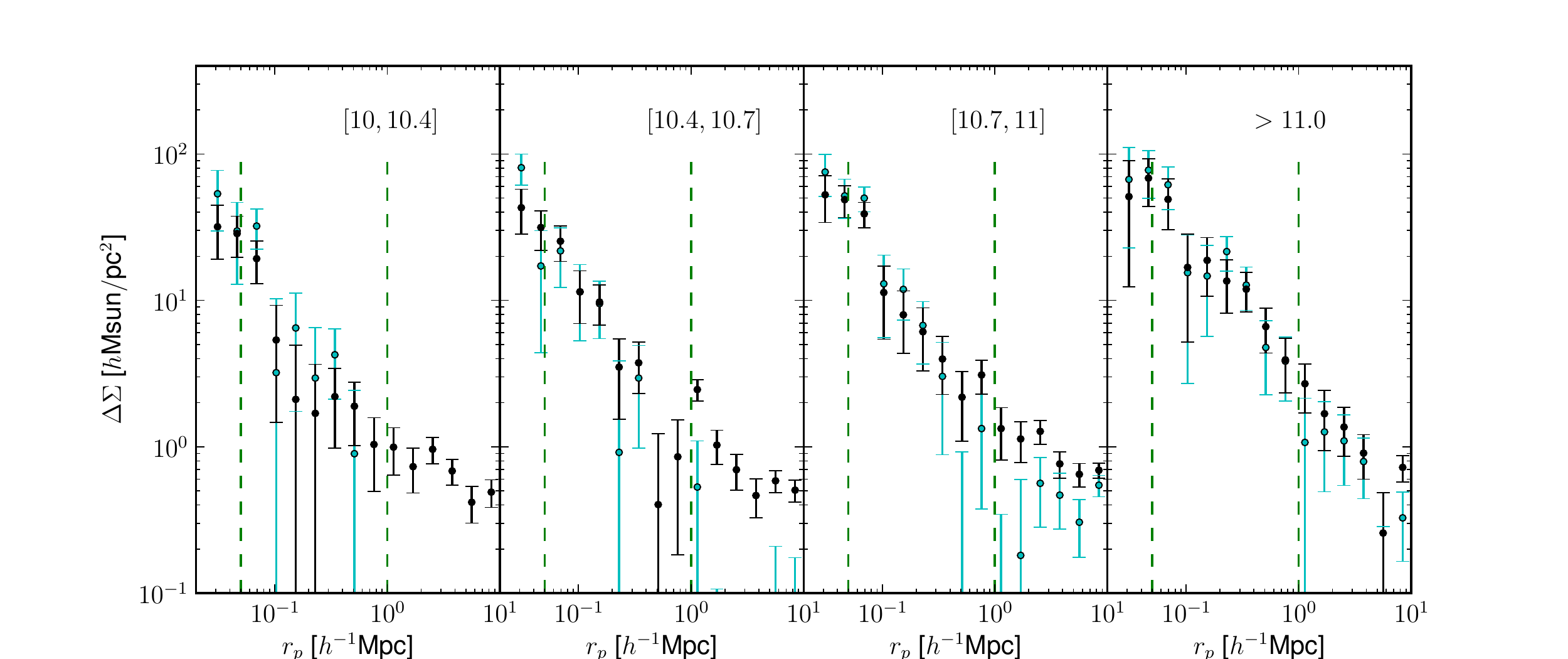}
\caption{\label{fig:lbgvsallblue}Lensing signals for blue Main sample galaxies vs.\ blue LBGs in
  stellar mass bins shown as \revy{black 
and light blue points} with errors, respectively.  The
vertical dashed \refresponse{lines show the lower and upper limits} in $r_p$ used for the fits of the LBG signals to NFW profiles.}
\end{center}
\end{figure*}

\begin{figure*}
\begin{center}
\includegraphics[width=6.4in,angle=0]{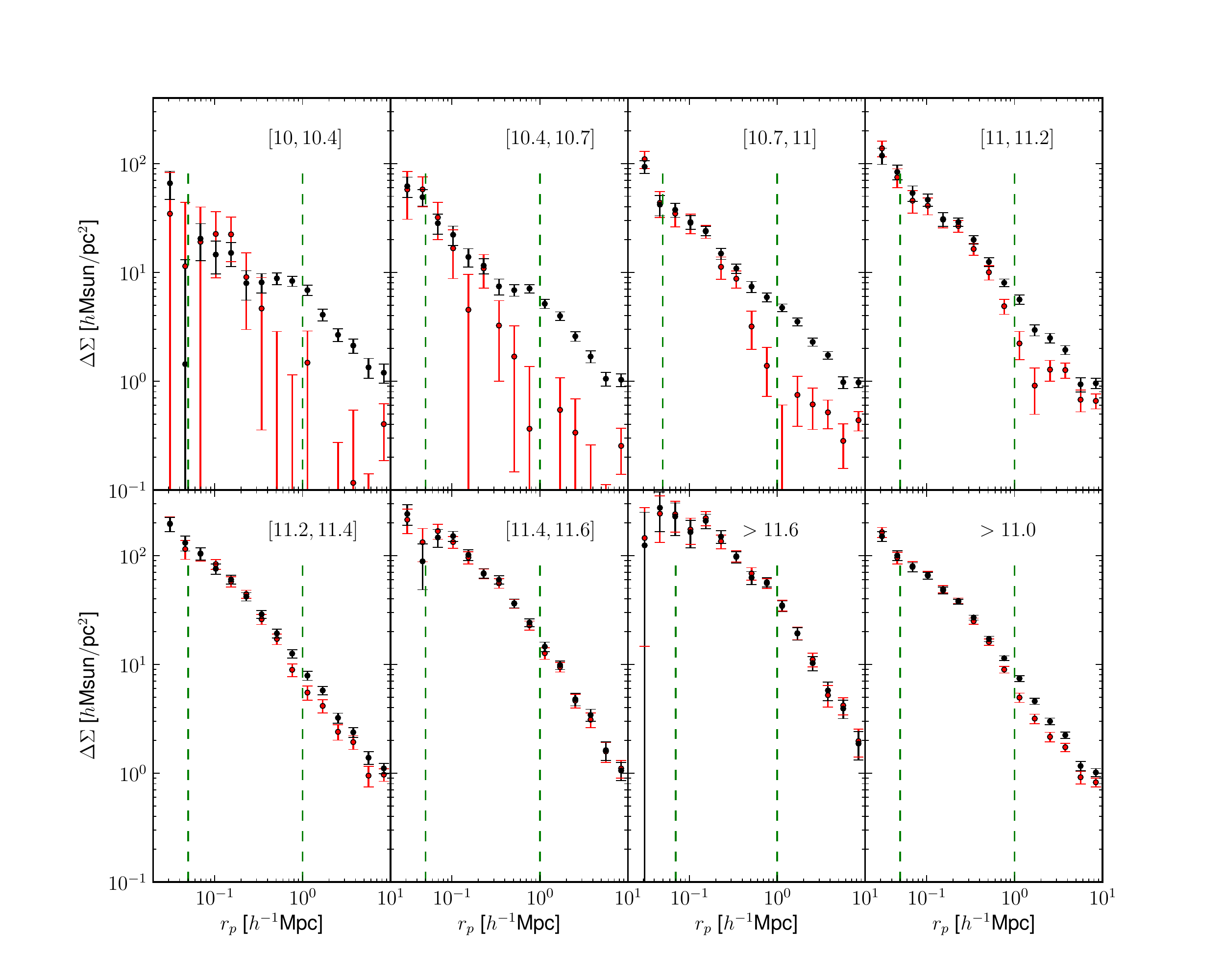}
\caption{\label{fig:lbgvsallred}Lensing signals for red Main sample galaxies vs.\ red LBGs in
  stellar mass bins shown as \revy{black
and red points} with errors, respectively.   The
vertical dashed \refresponse{lines show the lower and upper limits} in $r_p$ used for the fits of the LBG signals to NFW
profiles.}
\end{center}
\end{figure*}

Figure~\ref{fig:signals} shows the lensing signals for the LBGs in stellar mass bins, for red and
blue LBGs\footnote{\refresponse{The data shown in Figs.~\ref{fig:signals}--\ref{fig:lbgvsallred} can be downloaded in
  machine-readable format from \url{https://github.com/rmandelb/mandelbaum-data}.}}.  We first consider the basic question of consistency
between the signals for red and blue galaxies for $50\le R\le 1000$ \hkpc.  We carry out a
non-parametric $\chi^2$
test for consistency, defining
\beq
\chi^2 = \sum_i \frac{(\Delta\Sigma_\text{red} - \Delta\Sigma_\text{blue})^2}{\sigma_{\Delta\Sigma_\text{red}}^2 + \sigma_{\Delta\Sigma_\text{blue}}^2}
\eeq
where the summation is carried out over bins in $r_p$.  This test is interesting when the stellar
mass distribution for red and blue LBGs within the bin is sufficiently similar that 
differences in $\Delta\Sigma$ cannot come just from the different stellar mass distribution (this
condition is violated by the broad $M_*>10^{11}M_\odot$ bin, which is excluded from this test).  

Table~\ref{T:chi2} shows the $\chi^2$, degrees
of freedom, and $p$-values for each stellar mass bin and combined across all stellar mass bins.
The number of degrees of freedom appears larger than expected from the number of data points in
Fig.~\ref{fig:signals} because the data were rebinned in the figure to make it easier to read.  As
shown, adopting the convention that $p<0.05$ indicates a lack of consistency between the
red and blue samples, there is a clear inconsistency for three of the 
bins, and a marginal inconsistency ($p=0.06$) for a fourth bin.  In the other bins, the errorbars
are sufficiently large for the blue LBG sample that it is difficult to draw any conclusion.  
The $\chi^2$ over all seven bins is $p=1.2\times 10^{-4}$, ruling out the 
null hypothesis that red and blue galaxies at all stellar masses in this work have consistent
distributions of halo masses.
\begin{table}
\begin{center}
\begin{tabular}{llll}
\hline\hline
Stellar mass range & dof & $\chi^2$ & $p$-value \\
\hline
$[10,10.4]$ & 15 & 19.7 & 0.18 \\
$[10.4,10.7]$ & 15 & 18.8 & 0.22 \\
$[10.7,11]$ & 15 & 26.8 & 0.03 \\
$[11,11.2]$ & 15 & 24.2 & 0.06 \\
$[11.2,11.4]$ & 15 & 29.8 & 0.01 \\
$[11.4,11.6]$ & 15 & 28.7 & 0.02 \\
$>11.6$ & 15 & 18.8 & 0.22 \\
All & 105 & 166.9 & $1.2\times 10^{-4}$\\
\hline
\end{tabular}
\end{center}
\caption{Results of a $\chi^2$ test for consistency between the lensing signal for red and blue LBGs
  in each stellar mass bin, and across all stellar masses.  This test uses the signals from $50$--$1000$
  \hkpc.\label{T:chi2}}
\end{table}


As another diagnostic of the lensing profiles, we also consider the lensing profiles of all SDSS
Main sample galaxies using the same stellar mass and color divisions, but without the LBG selection
imposed.  This provides an alternative means (besides the mock catalogs used in
Sec.~\ref{sec:mocks}) of checking whether the LBG selection chooses \refresponse{a 
biased} subset of the central galaxies.  Ideally, the characteristic ``bump'' that
satellites can cause on $\sim 300$~\hkpc\ to $1$~\hmpc\ scales should disappear when comparing the
signal for all Main sample galaxies against that for LBGs, but the signal on smaller scales should
be essentially preserved.  This comparison is shown in Figs.~\ref{fig:lbgvsallblue}
and~\ref{fig:lbgvsallred} for blue and red LBGs, respectively.  Fig.~\ref{fig:lbgvsallblue} 
uses a single wide bin at high stellar mass due to the small number of blue galaxies above
$M_*>10^{11}~M_\odot$.  As shown, the signals for blue Main sample galaxies and LBGs in a given
stellar mass bin are nearly identical, with the LBGs showing some signal deficits above $\sim 300$~\hkpc\
as expected.  The differences are small  presumably because only a small fraction of blue
galaxies tend to be satellites in higher mass halos \citep[e.g.,][]{2006MNRAS.368..715M}.  For the
red galaxies, as shown in Fig.~\ref{fig:lbgvsallred}, the signals for Main sample red galaxies and
LBGs at fixed stellar mass are quite consistent below $\sim 300$~\hkpc, and show a characteristic
difference above that indicating that many Main sample red galaxies are satellites (especially at
lower stellar mass), and that the LBG
selection is effectively removing most or all of the satellites.  This result provides a separate
validation of the LBG criterion as a way of selecting a fair sample of central galaxies, in addition to the
mock catalog-based validation from Sec.~\ref{sec:mocks}.

\begin{figure*}
\begin{center}
\includegraphics[width=0.8\textwidth,angle=0]{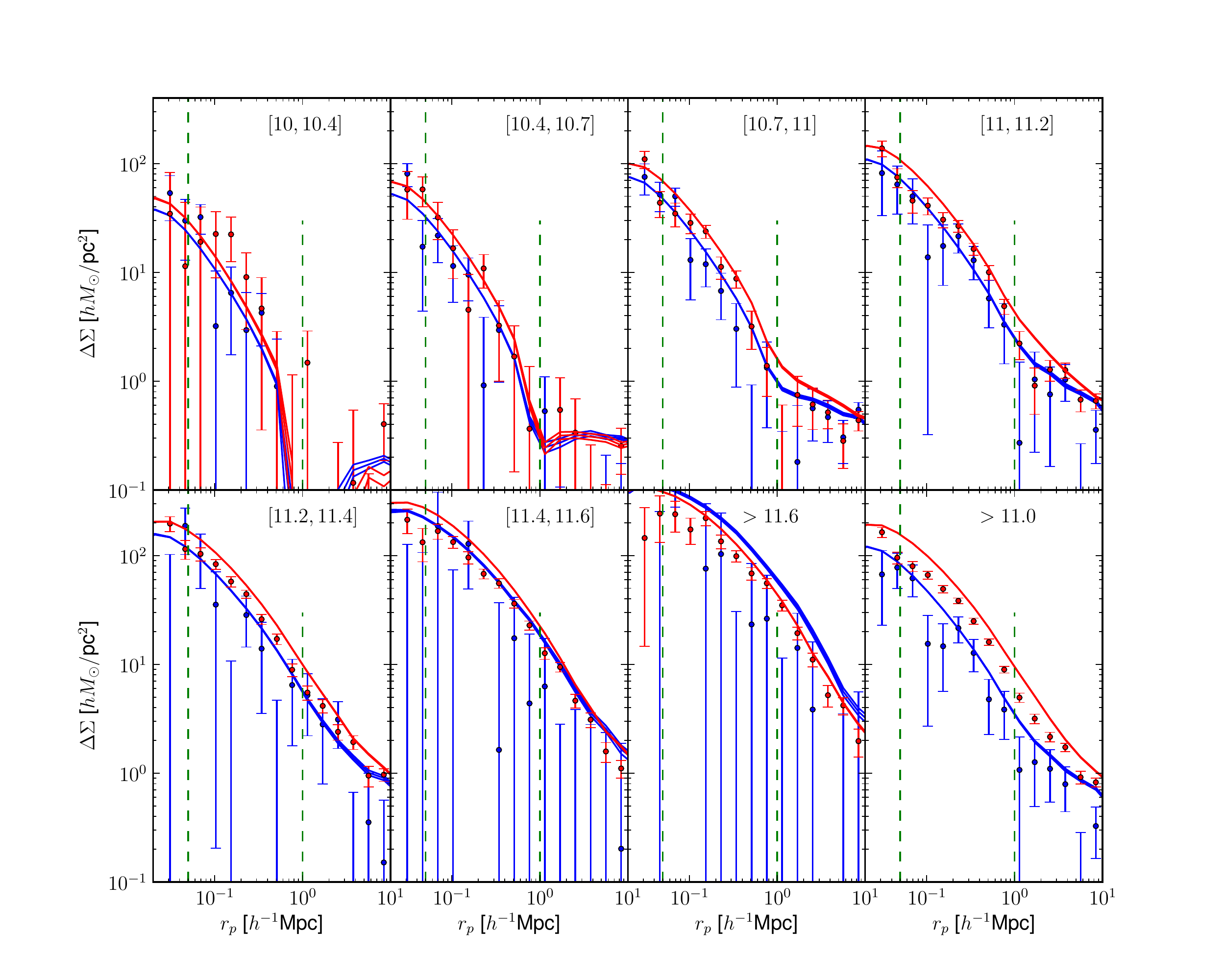}
\caption{\label{fig:comp_mocks_data} \refresponse{For all stellar mass bins as labeled on the plot, we compare
  the lensing $\Delta\Sigma$ profiles for red and blue LBGs in the data (the points with errors,
  same as \revsw{in} 
  Fig.~\ref{fig:signals}) with those in the mock catalogs (the solid lines).  The
vertical dashed \refresponse{lines show the lower and upper limits} in $r_p$ used for the fits of the LBG signals to NFW
profiles.}}
\end{center}
\end{figure*}

Finally, as a validation of our use of mock catalogs to correct for biases in the fitting
procedure, Fig.~\ref{fig:comp_mocks_data} shows, for \refresponse{all} stellar mass bins and both colors, a
comparison between the real \revsw{LBG $\Delta\Sigma$} and the LBG $\Delta\Sigma$ profiles in the
mock catalogs. \revall{These were computed via direct cross-correlation of the simulated LBGs and the
  (sub-sampled) dark matter density distribution in the simulations.}
It is important to keep in mind that these are not fits, and that the semianalytic model in the
mock catalogs was not specifically tuned \revsw{to fit lensing profiles at all.}  With that caveat in mind, the agreement between the
data and mock catalogs is quite remarkable \refresponse{especially for the bins with $M_* <
    10^{11.2}~M_\odot$}.  While the signal for blue galaxies \revall{in} the mocks is slightly high compared to that in the
data, the match is in general close enough that the mocks cannot have too different a halo mass distribution (as a function of
stellar mass and color) compared to the real data; for a more detailed exploration of this point,
see \revall{\cite{wenting-tmp}}. \refresponse{The possible exception to this statement is in the
  bins with $M_*>10^{11.4}~M_\odot$, in which there are almost no lenses in the real data, and our
  constraints are therefore extremely weak.} \revall{The red LBG data in the highest-mass bins show more tension, at roughly the
  20 per cent level (30 per cent in mass).}

\subsection{Mass estimates}\label{subsec:mass}

\begin{figure}
\begin{center}
\includegraphics[width=\columnwidth,angle=0]{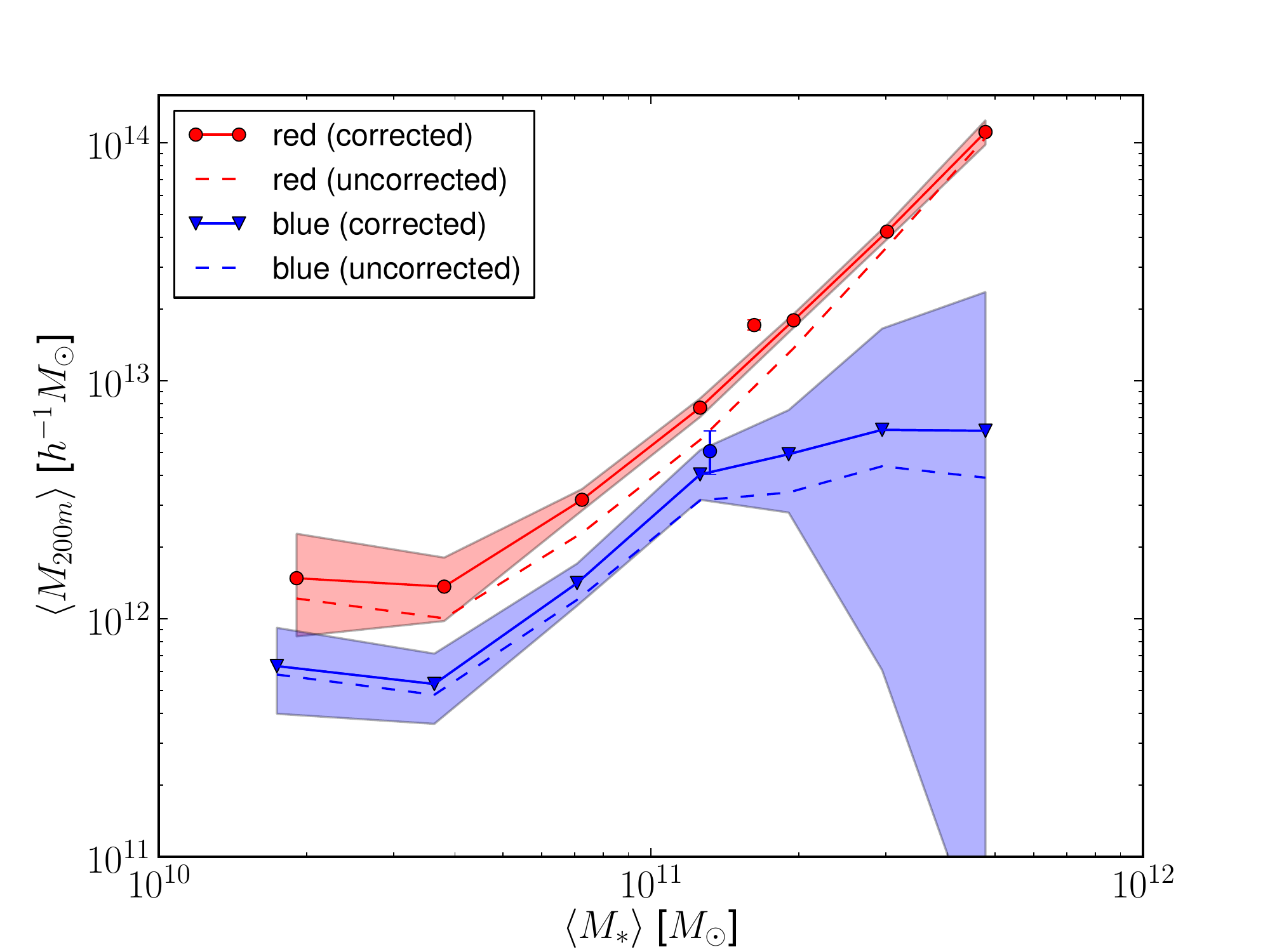}
\includegraphics[width=\columnwidth,angle=0]{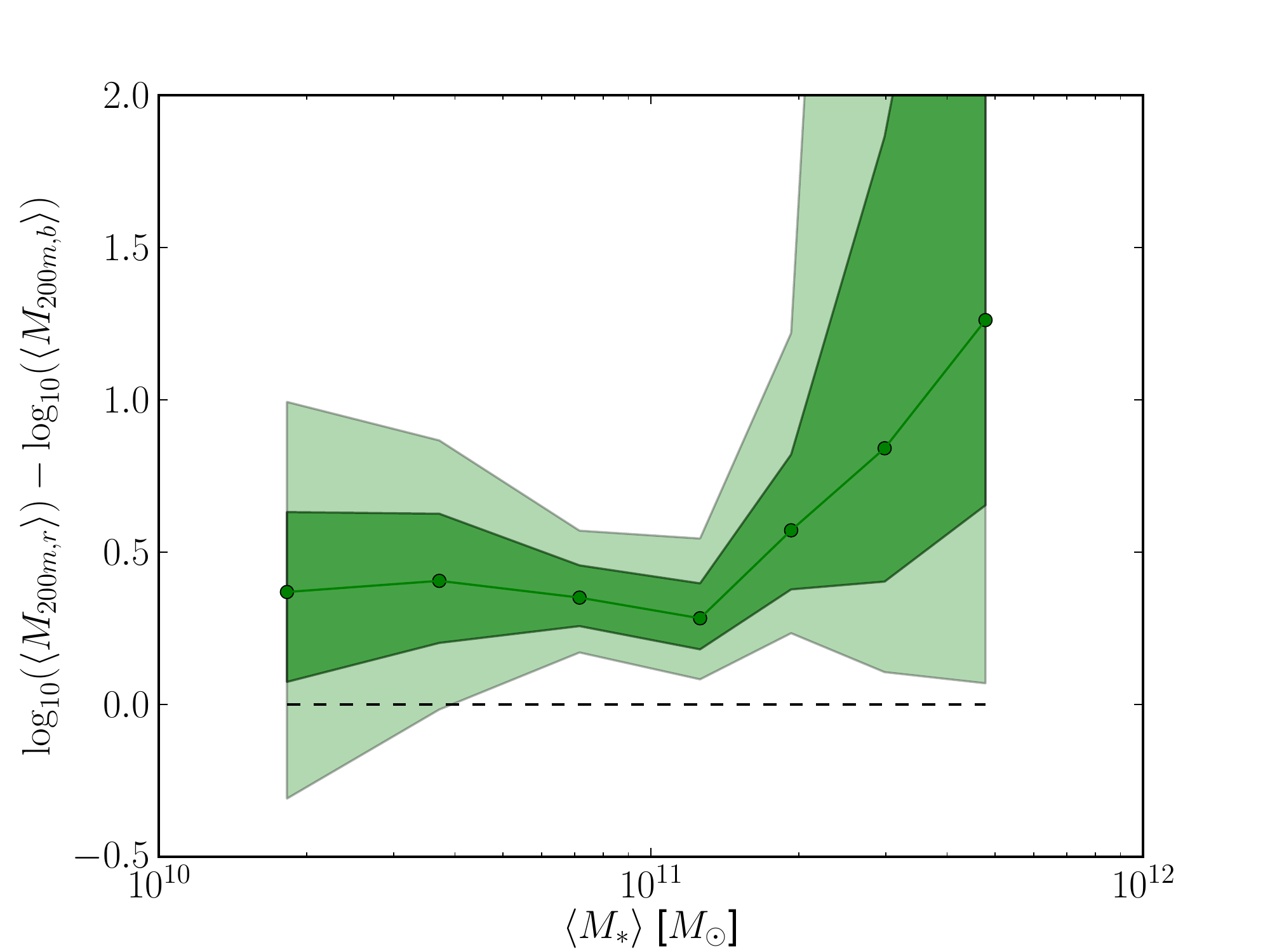}
\caption{\label{fig:mass} {\em Top:} The mean halo masses inferred from NFW fits to $\Delta\Sigma$
  as a function of the average stellar mass for each bin, shown separately for red (red
  circles) and blue (blue triangles) LBGs, along with shaded regions showing the 16--84 percentile
  range.  The dashed lines show results without the mock catalog-based corrections.  The single red
  and blue points shown separately with errorbars are the results for the single bin with
  $M_*>10^{11}M_\odot$, which has a very broad stellar mass distribution that is different for the
  two color samples.
{\em Bottom:} The difference in mass between the red and blue samples,
  expressed as $\log_{10}\langle M_{200m,\text{red}}\rangle-\log_{10}\langle M_{200m,\text{blue}}\rangle$.  The
  dark and light shaded regions show the 16--84 percentile and 2.5--97.5 percentile ranges,
  respectively.}
\end{center}
\end{figure}

In this section, we present mass estimates for the red and blue LBG samples.
Results are shown in Fig.~\ref{fig:mass}.  For the two lowest mass samples, the mass
difference (red $-$ blue) is consistent with 0 at approximately the $1.5\sigma$ level, while for the
higher stellar mass samples, the difference is positive at $>2\sigma$.

The results in Fig.~\ref{fig:mass} include the corrections from mock catalogs necessary to ensure
unbiased recovery of the average central halo masses, described in
Sec.~\ref{sec:mocks}.  In particular, we include two separate corrections: a
color- and stellar mass-dependent correction for biases in the fitting procedure in recovering
$\langle M_{200m}\rangle$ (due to basic inadequacy of the NFW fitting, plus satellite contamination;
Sec.~\ref{subsec:massfidelity}), and a correction for the fact that the central LBG sample has a
slightly different average mass compared to the full central galaxy sample
(Sec.~\ref{subsec:comppur}).  As shown previously, these corrections go in the same direction
for red and blue galaxies, with slightly different magnitude.  The top panel of
Fig.~\ref{fig:mass} also shows the results without the corrections; it is clear that these
corrections cause $>1\sigma$ changes in the mass estimates, but are not the reason for 
our conclusion that red and blue central galaxies have different average halo masses
in fixed stellar mass bins.

  The corrected values of mass and their uncertainties are tabulated in Appendix~\ref{app:mass}, Table~\ref{tab:mass}.

\subsection{Systematic error budget}\label{subsec:systematics}

Here we discuss the primary sources of systematic uncertainty in our results.  As discussed in
\cite{2013MNRAS.432.1544M}, a reasonable estimate of the multiplicative calibration uncertainty in
the lensing signal using this catalog is at most\footnote{This should be considered more like a
  $2\sigma$ range than a $1\sigma$ range.} 5 per cent, when considering shear systematics and
photometric redshift systematics together.  Since NFW mass fits scale like $\Delta\Sigma^{3/2}$ on
small scales, this implies an 8 per cent overall mass calibration uncertainty at the $2\sigma$ level, correlated
between all the samples (i.e., it would cause coherent shifts of all results in the same way).

However, Fig.~\ref{fig:mass} clearly reveals a more important source of systematic uncertainty.  As
shown, the correction factor that we apply to remove the effects of incompleteness, impurity, 
scatter in the mass distribution and other uncertainties in our modeling procedure is significantly
larger than the statistical error for nearly all the red samples, and corresponds to roughly
$1\sigma$ statistical errors for most of the blue samples.  Thus, the validity of this correction
factor, which encodes our understanding of multiple different issues as outlined in
Sec.~\ref{sec:mocks}, is our primary source of systematic uncertainty.  As already noted in
Sec.~\ref{subsec:lensing}, there are several empirical tests of the lensing signal that validate our
conclusions from the mock catalogs, so we have reason to believe the uncertainty is significantly smaller than the
size of the correction itself.

The corrections that were determined in Sec.~\ref{sec:mocks} are a combination
of the mass incompleteness correction, and a correction to our best-fit masses to estimate the mean
mass (given the halo mass distribution, which if broad, can result in significant differences
between best-fit and mean masses).  The results in \revall{\cite{wenting-tmp}} provide some
insight into the uncertainty in the second part of the correction factors.  In that work, multiple
different mock catalogs were used to model the LBG lensing and clustering signals (using the same
stellar mass bins, but without color division).  \revsw{The halo mass distribution were shifted by
  an overall multiplicative factor to match the lensing signals for LBGs in stellar mass bins, and
  the mean masses for all models (after this shift) were compared to get a scatter on the mean.
  This scatter in the mean mass reflects the different shapes of the halo mass distributions in the
  different 
  models.}  For the seven stellar mass bins, excluding the
broad one with $M_*>10^{11}~M_\odot$, the scatter in the base-10 logarithm of the mean halo masses across all the mock
catalogs is 0.035, 0.028, 0.069, 0.055, 0.060, 0.081, and 0.11 after correcting to our adopted halo
mass definition.  \revsw{Adopting these as systematic uncertainties is somewhat conservative, given
  that the extreme cases that drive the scatter to large values had halo mass distributions that may
  have resulted in an initial failure to reproduce LBG lensing signals.}

Table~\ref{tab:mass} shows that if we assume the same uncertainties for both colors, the ratio of
\revsw{(conservative) systematic to statistical uncertainty for red central LBG $\langle M_{200m}\rangle$ in our stellar
mass bins is $0.2$, $0.2$, $1.8$, $1.4$, $2$, $2.8$, and $2$.  For blue central LBGs, these ratios
(using the smaller upper errorbar) are $0.2$, $0.2$, $0.8$, $0.6$, $0.4$, $0.2$, $0.2$.}  Thus we
conclude that our conservatively-adopted systematic uncertainty is subdominant to statistical errors in all
cases for the blue LBGs, while for the red ones, they are comparable to the statistical errors above
$M_*>10^{10.7}~M_\odot$.  However, even in the \revsw{highly unlikely} case that the results shift by the entirety of
that systematic uncertainty in opposite directions for red and blue galaxies, this would not 
remove the majority of the observed mass difference between red and blue central galaxies.

\revsw{To assess the systematic uncertainty in the incompleteness corrections, which account for the
  fact that the mean halo masses of central LBGs differ from those of all central galaxies, our
  procedure is as follows.  We adopted two other semi-analytic models from
  \cite{2011MNRAS.413..101G} and \cite{2015MNRAS.451.2663H}, which are based on the WMAP1
  \citep{2003ApJS..148..175S} and Planck \citep{2014A&A...571A..16P} cosmologies, respectively.  The
  former has a similar semi-analytic prescription as the mock catalog used throughout this work,
  whereas the latter follows a different prescription (for more detailed comparison between the LBG
  samples in these mock catalogs, see \revall{\citealt{wenting-tmp}}).  These models were specifically
  chosen due to the fact that their halo mass distributions at fixed stellar mass are notably
  different from those in our fiducial mock catalog, resulting in a substantial mismatch when
  compared with the observed lensing signals
  (unlike for our fiducial mock catalog), and providing a quite
  conservative upper bound on systematic uncertainty in the incompleteness correction.  For these
  models, the completeness correction for blue (red) LBGs ranged from 1--4 per cent lower (5 per
  cent lower to 8 per cent higher) than our fiducial one, with some stellar mass dependence in both
  cases.} \refresponse{Fig.~\ref{fig:mass_compl} includes results from these alternate mock
  catalogs, in addition to our fiducial one.}

\revall{As discussed in Sec.~\ref{subsec:lbg}, the samples we use for our analysis are not
  volume-limited.  We carried out our NFW fitting procedure to signals calculated using only
  $z<0.08$ LBGs to quantitatively check what happens to the best-fitting masses. We found that with
  the blue $z<0.08$ LBG sample, there was a slight tendency for the NFW masses to decrease by $15$ per cent
  compared to our main results, while for the red $z<0.08$ LBG sample, the masses increase by
  typically 15 per cent.  Thus, use of a volume-limited sample is not driving our conclusions about
  mass differences; indeed, with a $z<0.08$ sample we would have found 30 per cent {\em larger} mass
differences, albeit with substantially larger statistical errors as well.}

Another source of systematic uncertainty is the assumed cosmology, which largely
affects the values of concentration used in the fitting.  We will consider this problem under the
more general
umbrella of concentration-related uncertainties, which also include any errors in the adopted $c(M)$
(even for the correct cosmology) and the impact of assembly bias (which could cause opposing errors
in the concentrations for blue and red galaxies: red ones should reside in older halos which are
more concentrated).  In principle, some errors in the $c(M)$ relation will be removed by the 
mock catalog-based correction factors.  For example, if the $c(M)$ relation 
used for the fits is wrong compared to both reality and the mocks, then our final results will be 
unaffected after applying the mock catalog-based corrections.  Since our primary conclusion 
is that red and blue central galaxies reside in halos with different masses on average, 
we focus particularly on the following
question: could the impact of assembly bias on the concentrations of halos hosting red and blue 
central galaxies be causing us to incorrectly draw our primary conclusion, rather than it being a
real effect?

If red galaxies actually live in halos with higher concentrations than we assumed and than in
the simulations, and vice versa for blue galaxies, then the inferred halo masses for red (blue)
galaxies are too high (low).  The sign of this effect is
therefore correct to explain our result, 
but it is unclear whether the magnitude could be as large as the observed halo mass differences
(factors of 2--2.5).  To address this question, we note that \cite{2002ApJ...568...52W} found a
scatter in $\log_{10}{c}$ of $\sim 0.12$.  Taking the extreme assumption that red and blue central
galaxies \revall{at fixed halo mass represent halo samples split into the upper and lower part of
the $c(M)$ distribution depending on the red galaxy fraction, $\langle
c_\text{red}\rangle$ and $\langle
c_\text{blue}\rangle$ can differ by up to $\sim 35$ per cent (depending on the mass range and red
fraction, considering reasonable values of both quantities). Redoing 
the fits with 
modifications in concentration that are of that order for the stellar mass bin from $10^{10.7} \le
M_*/M_\odot < 10^{11}$ causes the mass ratio to change by about 12 per cent.}  Even with extreme assumptions,
therefore, assembly bias cannot be responsible for any sizeable part of the observed factor of 2--2.5
difference in mass between red and blue galaxies\footnote{Also, for red galaxies, increasing
  the concentration by 35 per cent results in notably worse $\chi^2$ values; the data in fact prefer
  {\em reduced} concentrations rather than increased ones.  For blue galaxies, there is less discriminating power between concentrations due to
  the lower signal to noise.}.  Under less extreme assumptions, there
may be mass uncertainties at the few per cent level due to assembly bias affecting 
the halo concentrations for red and blue central galaxies.  This is subdominant to the overall
modeling uncertainties discussed above.

\revall{As a final test of concentration-related uncertainties, we reran the fits with the
  concentration free but with a lognormal prior about the adopted $c(M)$ relation used for the rest
  of this work.  In general, the concentrations went down, and the masses went up.  For red LBGs, we
  find that the masses typically increase by $\sim 20$ per cent in the data, and $\sim 11$ per cent in the mocks.
  Thus, when we consider the change in the mass correction factor that is based on the fits to the
  mock catalogs, our estimated red central galaxy halo masses would go up by typically $\sim 9$ per cent (but are also more noisy due to the
  extra degree of freedom).  For blue LBGs, we find that the
  masses typically increase by $\sim 30$ per cent in the data, and $\sim 20$ per cent in the mocks.  Thus, the net increase
in the estimated blue central galaxy halo masses when taking into account the changed correction
factor would be $\sim 10$ per cent.
Thus, our estimates for both blue and red central halo masses would increase by $\sim 10$ per cent,
leaving the mass ratio unchanged on average, but noisier.  The increase in noise and the lack of
change in our conclusion about halo mass bimodality justifies our choice to keep the
fixed-concentration results as our primary ones.}

\subsection{Comparison with previous results}\label{subsec:comparison}

In this section, we compare our results with previous estimates of the galaxy type-dependent
relationship between stellar and halo mass of
central galaxies.  In all
cases when comparing with papers that have multiple redshift ranges, we use the closest available
range.  Stellar masses have been converted to our adopted units and value of $h$ and IMF, while halo
masses have been converted to our adopted convention, $M_{200m}$ with our value of $\Omega_m$.
\revall{We have not accounted for more subtle differences in halo vs.\ stellar mass relations that
  could arise from, e.g., different assumptions about halo concentration vs.\ mass relations.}  \revsw{Differences in stellar mass modeling can lead to additional (non-IMF-related) systematic offsets between
  stellar mass estimates that can be significant
\citep{2012ApJ...746...95L}, but are difficult to estimate and correct}.   When
comparing with previous \revsw{SDSS} results, we use the MPA/JHU stellar mass estimates \revsw{that
  were adopted by many of those works}.  The exception to this is
\cite{2011MNRAS.410..210M}, who used stellar masses from the \cite{2001ApJ...550..212B} approach;
however, the fitting formulae in \cite{2009MNRAS.398.2177L} show that for the range of stellar
masses used in this work, the MPA/JHU and \cite{2001ApJ...550..212B} agree on average, so we do not apply
a correction.  In all cases, the plotted quantity is $\langle M_{200m}\rangle$ at
fixed stellar mass, either in bins in stellar mass (shown using points) or from a 
parametric model or continuous function based on finely binned halos in simulations (shown using
curves).  The errorbars indicate error on the mean, rather than intrinsic scatter about the mean.

\begin{figure*}
\begin{center}
\includegraphics[width=\textwidth,angle=0]{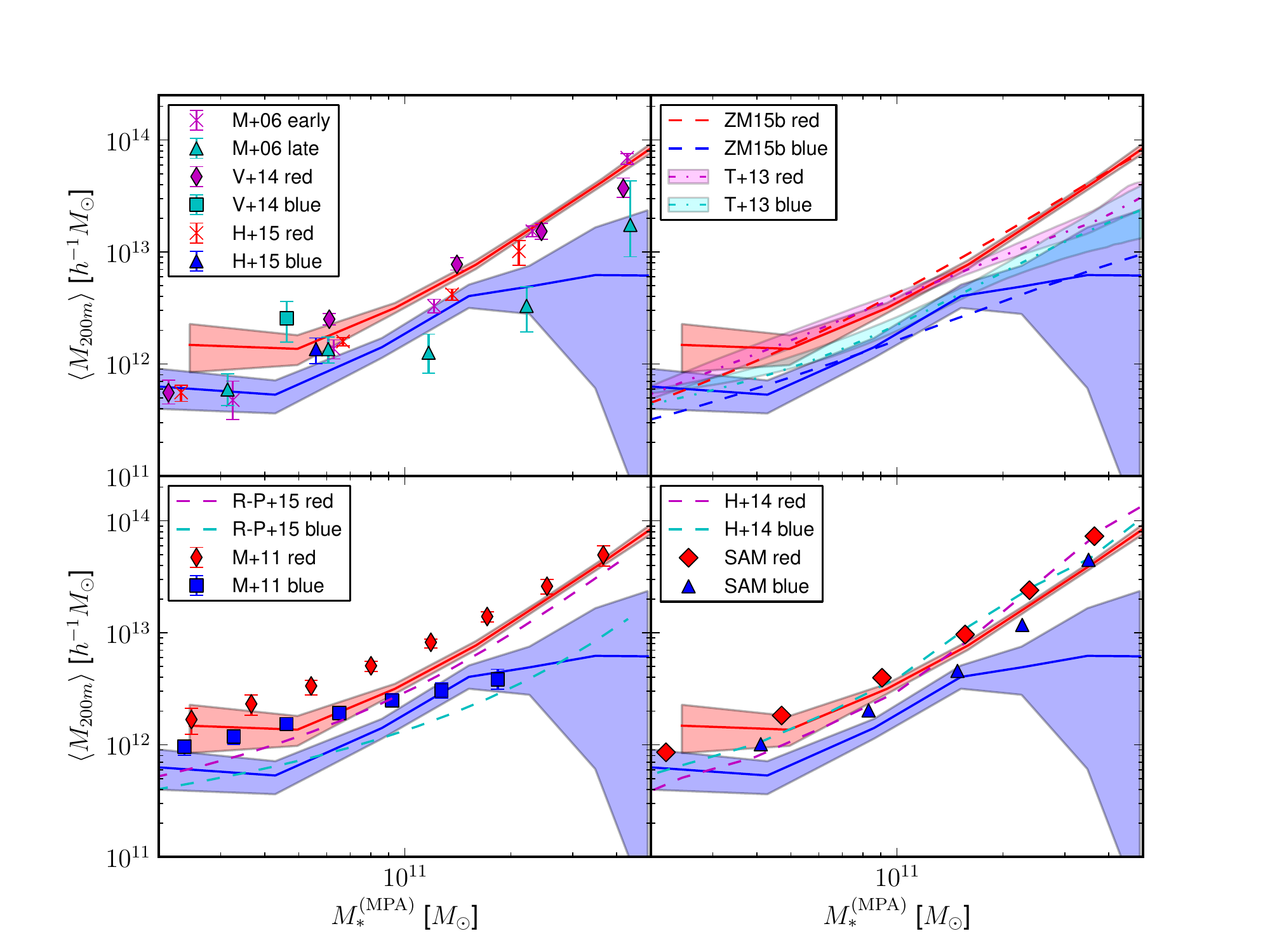}
\caption{\label{fig:compare} All panels show the mean central halo mass as a function of stellar mass
  for red and blue galaxies from this paper (Fig.~\ref{fig:mass}) as shaded regions.  The top left
  panel shows results from measurements of galaxy-galaxy lensing by 
  \protect\cite{2006MNRAS.368..715M}, \protect\cite{2014MNRAS.437.2111V}, and
  \protect\cite{2015MNRAS.447..298H}; the bottom left panel shows results from
  \protect\cite{2011MNRAS.410..210M} (satellite kinematics) and \protect\cite{2015ApJ...799..130R}
  (HOD interpretation of abundance and clustering); the top right panel shows results from HOD
  modeling of lensing
  and clustering measurements by \revall{\protect\cite{2015arXiv150906758Z}} and
  \protect\cite{2013ApJ...778...93T}; and the bottom right panel shows age-matching mocks from
  \protect\cite{2014MNRAS.444..729H} as well as the semi-analytic catalogs used throughout this work..  All
  errors are $1\sigma$.}
\end{center}
\end{figure*}

A number of previous works measured the color-dependent average
halo mass for central galaxies in bins in
stellar mass.   \cite{2006MNRAS.368..715M},
\cite{2014MNRAS.437.2111V}, and \cite{2015MNRAS.447..298H} measured galaxy-galaxy lensing by
galaxies split into passive and active categories as a function of stellar mass, and used a simplified halo
model to interpret the results in terms of typical central halo masses (along with a satellite
term).  The top left panel of Fig.~\ref{fig:compare} shows a comparison between our results (shown as
shaded regions that are the same as those in Fig.~\ref{fig:mass}) and those three works.  As shown,
the results of \cite{2006MNRAS.368..715M} are largely consistent with ours, with a hint of slightly
lower masses for red galaxies below $M_*^{\text{(MPA)}}=10^{11}~M_\odot$.  However, this could be a
consequence of modeling differences (since a simple halo model was used to separate out
satellite and central contributions, which are quite important for low-mass red galaxies).
Alternatively, this is the only work against which we compare that used a morphological- rather than
color-based separator (indicated on the plot using ``early/late'' instead of ``red/blue'');
while morphology and color are correlated, they do not split up the data in exactly the same way
and could drive some of the observed differences.  Also
note that our current results have a slightly higher $S/N$ than the earlier \cite{2006MNRAS.368..715M}
results, due to the larger area used and the better 
photometric redshifts for the source galaxies (which make our weighting closer to optimal).  This
$S/N$ difference, combined with the slight shift for the early-type galaxies, is the reason
that we now see a more statistically significant difference between the average halo
masses for red and blue central galaxies even below $M_*=10^{11}~M_\odot$.  The
\cite{2014MNRAS.437.2111V} results seem consistent with ours for red galaxies, but somewhat high for
blue galaxies. The \cite{2015MNRAS.447..298H} results are slightly lower than ours for red galaxies,
but agree for blue galaxies.   Differences in modeling and/or in stellar mass estimates may be at least
partly responsible for these differences, which in any case are not very statistically 
significant.

The lower left panel of Fig.~\ref{fig:compare} shows a comparison with two rather different results.
\cite{2011MNRAS.410..210M} used satellite kinematics of central galaxies to infer the relation
between halo and stellar mass.  Their results agree with ours at high mass, but are significantly
higher for $M_*^{\text{(MPA)}}<10^{11}~M_\odot$ (however, at the lowest masses the discrepancy is
not significant, primarily due to our much larger errors).  One possible explanation for a discrepancy
of this sign is satellite contamination of their central galaxy sample
\citep{2011MNRAS.410..417S,2015arXiv150407632L}, which could be more important below
$M_*\sim 10^{11}M_\odot$; however, \cite{2011MNRAS.410..210M} applied additional isolation criteria
to avoid this problem.  There are slight differences in their sample selection due to their
different isolation criterion and color cuts, though it is unclear why those would cause a
difference that always goes in the same direction.  The most likely remaining explanations are
modeling issues that may be more 
relevant at lower mass, where the number of satellites per halo becomes quite small on average; we
defer further exploration to future work.  \refresponse{For a comparison of several kinematics-based
stellar vs.\ halo mass relations, see \cite{2013MNRAS.428.2407W}.}

Also in the lower panel of Fig.~\ref{fig:compare} are 
the results of \cite{2015ApJ...799..130R}, who used a \revall{halo occupation distribution (HOD)} to interpret the 
stellar mass function and galaxy clustering of low-redshift galaxies in the SDSS.  We show their
mean halo mass as a function of stellar mass without errorbars, since the errors are very small
compared to ours.  Their results agree 
fairly well with ours for blue galaxies, but are below our results 
for red galaxies above $10^{11}~M_\odot$.  It is possible that in the absence of lensing, their halo
masses are more reliant on assumptions within their halo model to interpret the clustering.  
\revall{\cite{wenting-tmp}} show that even with the same cosmological model and
stellar mass function, the clustering signal can vary significantly depending on model assumptions.
We propose that this may be the primary cause for the discrepancy.

The top right panel of Fig.~\ref{fig:compare} includes two papers that interpreted galaxy-galaxy
lensing and clustering measurements jointly.  The results of \revall{\cite{2015arXiv150906758Z}} were obtained by applying a halo model to the galaxy-galaxy lensing and clustering
measurements of Main sample galaxies divided by color and stellar mass (i.e., data similar to the
``Main sample'' curves in Figs.~\ref{fig:lbgvsallblue} and~\ref{fig:lbgvsallred}).  As shown, these
agree extremely well with our LBG results, despite the significantly more complex modeling needed to
interpret those results.  The halo model used there must account for the contributions by those Main
sample galaxies that are satellites, and includes the effects of scatter between stellar and halo
mass.  The fact that our results are consistent with theirs after using our mock-catalog-based correction for that
scatter in the LBG signals is a non-trivial result. 

 The curves from \cite{2013ApJ...778...93T} in the top right panel of Fig.~\ref{fig:compare} were
obtained by jointly modeling the stellar mass function, lensing, and clustering in the COSMOS survey
using a formalism that essentially doubles the number of halo model parameters in order to
separately describe the results for red and blue galaxies.  As shown, for blue galaxies their
results are consistently slightly above ours, but with significant overlap of the error regions.
For red galaxies, their results are consistent with ours for $M_*<10^{11}~M_\odot$, but deviate to
significantly lower halo masses at high stellar mass.  As a consequence of these two differences,
the red vs.\ blue halo mass difference inferred from their results is smaller than ours.  At the
higher mass end, it is likely that their results are dominated by the higher signal-to-noise stellar
mass function and clustering constraints, which may be more dependent on modeling assumptions as
mentioned above in the discussion of the \cite{2015ApJ...799..130R} results.  \revy{It is also worth
noting that the mean redshift for this result is $0.36$, so there may be some small 
evolutionary effects complicating the comparison between their results and ours.}

Finally, the bottom right panel of Fig.~\ref{fig:compare} shows the results of two mock catalogs.
The first are the age-matching \citep{2013MNRAS.435.1313H} mock catalogs publicly released with
\cite{2014MNRAS.444..729H}, who used stellar mass rather than luminosity as in the original
age-matching paper.  We show the mean halo mass at fixed stellar mass without errorbars, since the error on the mean is
quite small. As shown, the $\langle M_{200m}\rangle$ curve for red central galaxies is slightly
steeper than in our LBG results.  However, as a basic consequence
of their age-matching procedure, their $\langle M_{200m}\rangle$ for blue
central galaxies are above the red ones for much of the stellar mass range.  This results in a
highly significant discrepancy with our observations of blue central galaxy halo masses.  
The inversion of the red and blue curves over a wide range of stellar mass, which results from the
assumptions of the method and some particular implementation details in \cite{2014MNRAS.444..729H},
is discussed in more detail in \revall{\cite{2015arXiv150906758Z}}.  It is worth noting,
however, that \cite{2014MNRAS.444..729H} explored their ability to predict observable quantities
associated with mixed galaxy samples containing central
and satellite galaxies.  In principle, a cancellation between effects seen in the two samples could
be removed when exploring a sample of purely central galaxies, giving a more severe discrepancy with
observations in our work compared to the moderate tension shown in the plots in
\cite{2014MNRAS.444..729H}. \revall{ As mentioned in Sec.~\ref{subsec:lbg}, the fact that our sample is not
volume-limited (and the one used by \citealt{2014MNRAS.444..729H} is volume-limited) cannot be responsible for these
differences.  A direct comparison of the $\Delta\Sigma$ profiles for all Main sample
  galaxies, and for LBGs, against predictions from the \cite{2014MNRAS.444..729H} is shown in
  Appendix~\ref{app:age-match}.}

The points shown in the bottom right panel of Fig.~\ref{fig:compare} are from the central galaxies
in the semi-analytic mock
catalogs used throughout this work (Sec.~\ref{sec:mocks}), using the same set of stellar mass bins
as for the data.  The statistical errors on the mean masses are not shown because they are very
small; the systematic uncertainties are larger (Sec.~\ref{subsec:systematics}).  As shown, the
curves for red central galaxies in the SAM-based catalogs are slightly above our observations, but
not in a significant way \revfix{except for \refresponse{at $M_*>10^{11.2}~M_\odot$, also reflected in Fig.~\ref{fig:comp_mocks_data}}}.  The deviation between our observations and the blue galaxy halo masses in
the mocks is \revfix{also more significant } at the highest stellar masses
($>10^{11}~M_\odot$) where we have very few galaxies.   Overall, the mock catalogs exhibit a difference
between the average halo masses for red and blue central galaxies in these bins that is consistently
of the right sign and slightly smaller magnitude compared to the real data. \revfinal{ This
  is remarkable since no clustering or lensing information
was used when tuning the parameters controlling formation and
quenching processes in this simulation.}

From a theoretical standpoint, it is interesting to consider how our results can be used.  For
example, the upper right panel of Fig.~\ref{fig:compare} shows a comparison against a halo model that
includes a prescription for quenching (the shutdown of star formation activity that transforms a galaxy
from active to passive) from \revall{\cite{2015arXiv150906758Z}} and
matches our results very well.  This quenching model, necessary to make a color-dependent HOD, adds
four parameters to the halo model, which is much fewer than the most obvious way to make a
color-dependent HOD (which would double the number of parameters).  
Different models for the quenching process -- i.e., different dependencies on stellar and/or halo
mass -- modify the mean halo mass at fixed stellar mass,
suggesting that one can use our central galaxy constraints from this paper to constrain quenching
models.  In contrast, the poor match to the average halo masses in the age-matching mocks from
\cite{2014MNRAS.444..729H} calls into question the assumptions underlying the galaxy-dark matter connection in
those mock catalogs.

\section{Discussion and conclusions}\label{sec:discussion}

In this work, we have used the galaxy-galaxy lensing signals of a sample of LBGs in the SDSS Main
galaxy sample to measure the mean dark matter halo masses for central galaxies as a function of
their stellar mass and color.  To validate the relatively simple interpretation of these lensing
signals in terms of average halo masses, we have used both simulation-based mock catalogs that closely
reproduce the observed properties of Main sample galaxies and LBGs, and cross-checks within
the observational dataset itself.  We observe a clear, $>3\sigma$ difference in the average halo masses of blue and
red central galaxies at fixed stellar mass, for all stellar masses above $6\times
10^{10}~M_\odot$.  The halo masses of the red (passive) central galaxies are higher than those
of blue (active) central galaxies. \revy{We have demonstrated that selection effects cannot be responsible
for this result, using a semi-analytic mock catalog that reproduces many properties of the
LBG sample.}

\revy{Our result that there is a bimodality in halo masses of central galaxies at fixed
stellar mass} can be related to the underlying physical
processes such as feedback.  For example, \cite{2012MNRAS.424.2574W} used indirect measures of halo
mass (the satellite counts around central galaxies) to infer a difference in average halo mass at
fixed stellar mass for isolated red and blue primary galaxies.  Their finding in a mock catalog that
exhibited the similar effect \citep{2011MNRAS.413..101G} was that this could be attributed to AGN
feedback suppressing star formation in central galaxies that have massive black holes, thus lowering
the average stellar mass for red galaxies at fixed halo mass (or, at fixed stellar mass, giving a
larger average halo mass for red primaries than for blue ones).

As discussed in Sec.~\ref{subsec:comparison}, our results are largely consistent with previous
lensing results \citep{2006MNRAS.368..715M,2014MNRAS.437.2111V,2015MNRAS.447..298H}, with only minor
tensions that may result from different modeling processes, stellar mass estimates, or ways of
dividing the sample into blue and red subsets.  We observe a clear tension with the results from
satellite kinematics \citep{2011MNRAS.410..210M} below $M_*=10^{11}~M_\odot$, likely due to modeling differences.
Our tension goes in the same direction for both the red and blue samples, with the kinematics
results giving a higher mass; thus, our work and theirs come to the same basic conclusion  that red galaxy samples have higher average
halo masses than blue galaxy samples at fixed stellar masses.
Our results for blue galaxies are consistent with those of \cite{2015ApJ...799..130R}, who used an HOD
model to interpret galaxy clustering and the stellar mass function, but our results for red galaxies
are in disagreement with theirs, likely due to the increased sensitivity to modeling assumptions
when using clustering to infer halo masses.  However, our results
are consistent \revall{with} those based on HOD modeling of galaxy-galaxy lensing and clustering by 
\revall{\cite{2015arXiv150906758Z}} using a formalism that incorporates the results of halo quenching.  We also observe a highly significant disagreement
with the blue central galaxy halo masses in the age-matching mocks from \cite{2014MNRAS.444..729H},
due to an inversion in those mocks of the average halo masses for red and blue central galaxies.

Our results clearly imply that high-fidelity mock catalogs that reproduce the galaxy-matter
connection cannot be produced using methods that neglect type-dependence, such as 
simple halo modeling.  In addition, the age-matching approach appears to give
average halo masses 
for blue central galaxies that are highly discrepant with our observations, as discussed in more
depth in Sec.~\ref{subsec:comparison}.  More sophisticated
methods that include a strong galaxy type-dependence in $\langle
M_{200m}\rangle$ (enforcing the correct sign of the difference) at fixed stellar mass seem to be necessary to
reproduce observations; as we have demonstrated, this can be achieved with semi-analytic models
or a type-dependent halo model.

\section*{Acknowledgments}

\refresponse{The authors thank Jeremy Tinker for providing the curves from
  \cite{2013ApJ...778...93T} shown in Fig.~\ref{fig:compare}, and the anonymous referee for
  suggestions that improved the presentation of results in this work.}
RM and YZ acknowledge the support of the Alfred P. Sloan Foundation.
WW acknowledges a Durham Junior Research Fellowship (RF040353). 
\revall{BH and SW were supported in part by the Advanced Grant 246797 ``GALFORMOD''
from the European Research Council, and BH by a Zwicky fellowship.  SM is supported by World Premier International Research Center
Initiative (WPI Initiative), MEXT, Japan, by the FIRST program `Subaru
Measurements of Images and Redshifts (SuMIRe)', CSTP, Japan, and by
Grant- in-Aid for Scientific Research from the JSPS Promotion of
Science (No. 15K17600).}

\bibliographystyle{mnras}

\bibliography{ms,newrefs,cosmo,simground}

\appendix

\section{\refresponse{Tests of the non-volume-limited nature of the sample}}\label{app:vollim}

\begin{figure*}
\begin{center}
\includegraphics[width=0.8\textwidth,angle=0]{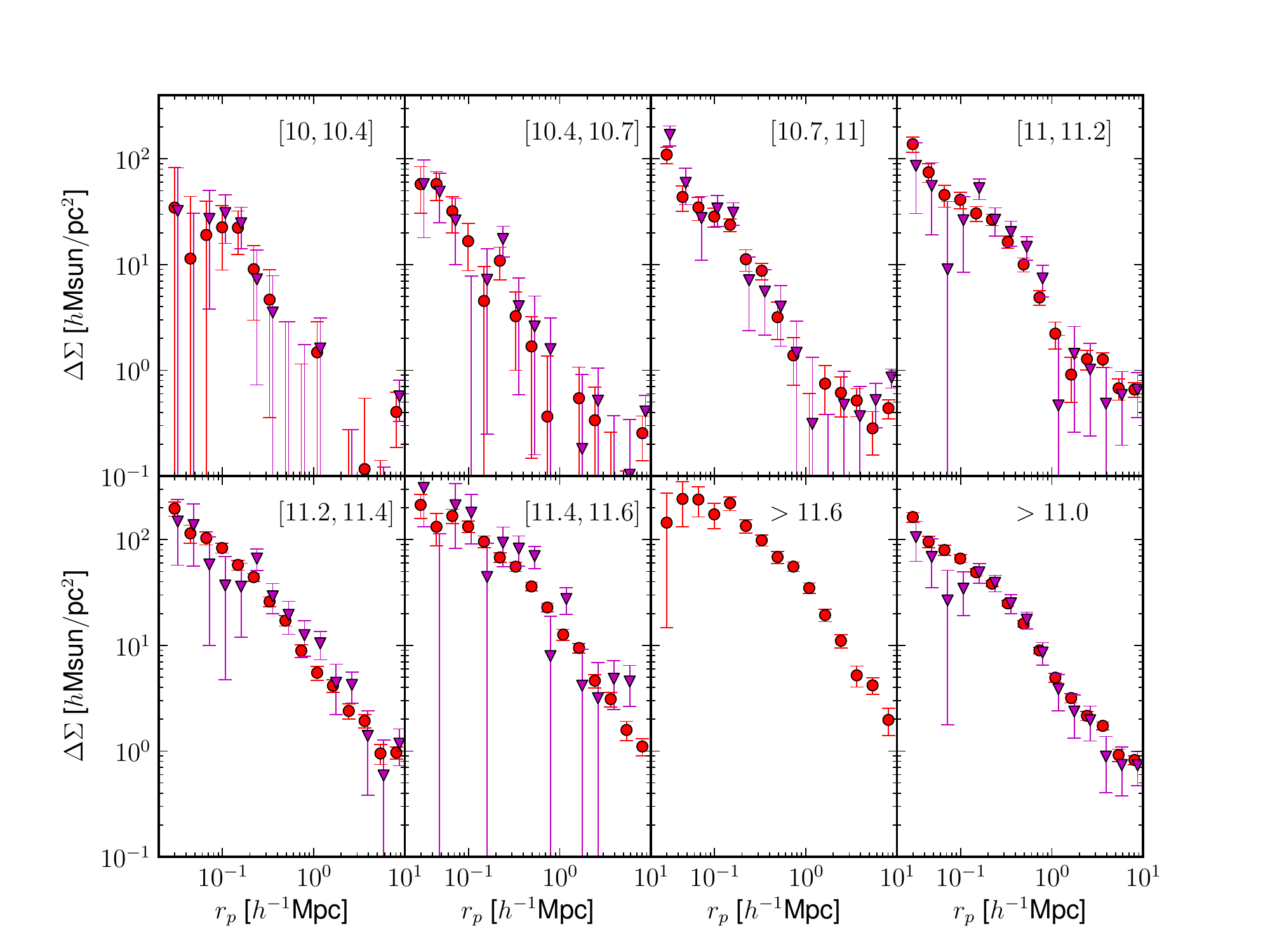}
\includegraphics[width=0.8\textwidth,angle=0]{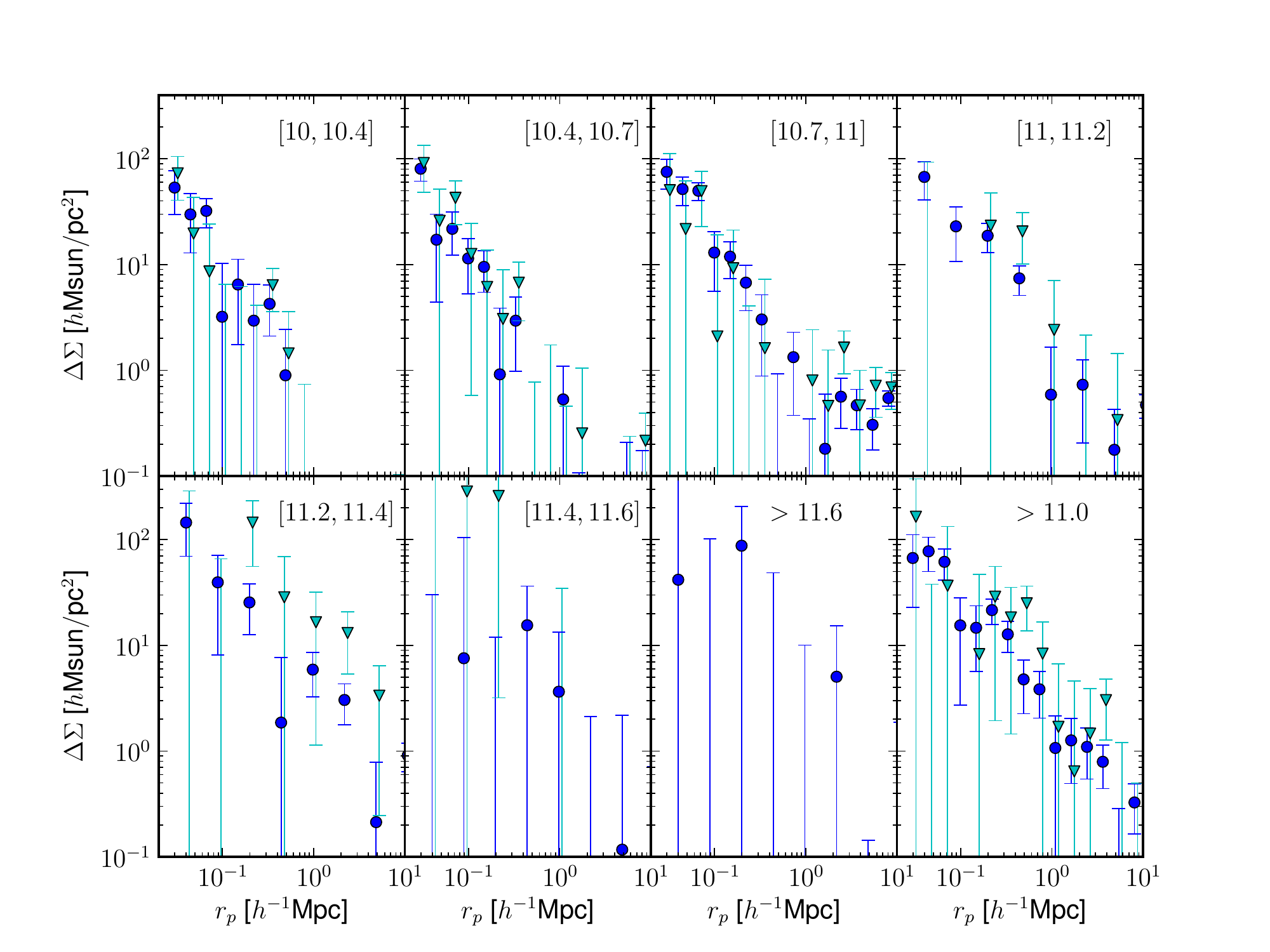}
\caption{\label{fig:vollim} \refresponse{Comparison between lensing signals for all LBGs and for
    $z<0.08$ LBGs for the red (top) and blue (bottom) subsamples, split into the same stellar mass
    bins as in the rest of this paper.  The data for all LBGs is the same as in
    Fig.~\ref{fig:signals}.  In both panels, the circles (triangles) show the results for LBGs
    ($z<0.08$ LBGs).  Neither panel has $z<0.08$ LBG results for the highest stellar mass sample,
    just because there are almost no lenses in that sample below that redshift.}}
\end{center}
\end{figure*}

\refresponse{As described in Sec.~\ref{subsec:lbg}, we compared the lensing signals for the full LBG
samples (as shown in Fig.~\ref{fig:signals}) with those for the LBGs at $z<0.08$.  While technically
the redshift below which the samples are volume limited depends on the stellar mass range, this
serves as an approximate test for how much it matters that the samples are not volume-limited.
Fig.~\ref{fig:vollim} compares the lensing signals for the red (top) and blue (bottom) LBGs with
those for $z<0.08$ LBGs.  The results are fully consistent with each other, supporting our choice
to proceed with the full LBG samples.}

\section{Mass results}\label{app:mass}

In Table~\ref{tab:mass}, we summarize our average halo mass results and other relevant statistics
\revsw{for red (top section) and blue (bottom section) central galaxies}.
The errorbars on the mean halo masses are actually the 16th and 84th percentile values of the
non-Gaussian error distribution.  We have used those same percentile values to give non-Gaussian
16th and 84th percentile values for the ratio of stellar-to-halo mass (normalized by
$\Omega_b/\Omega_m$).  Finally, the total corrections applied to the NFW masses, including the
effects of completeness and impurity of the LBG sample as well as systematics in the fitting
(e.g.\ due to the finite width of the halo mass distribution) are shown as the last column.  These
are the product of the two correction factors derived in Sec.~\ref{sec:mocks}.
\begin{table*}
  \begin{center}
    \caption{\label{tab:mass}Summary of our results for the average halo mass for central galaxies
      in fixed stellar mass bins \revsw{for red (top section) and blue (bottom section) central galaxies}.  The quantities that are tabulated are the lensing-weighted
      stellar mass of the galaxies taking into account their weight in the lensing
      measurement for our canonical (VAGC) stellar masses as well as for the MPA/JHU stellar masses, the
      mean halo mass, the ratio of central stellar mass to host halo mass normalized by $\Omega_b/\Omega_m$ assuming a
      \protect\cite{2014A&A...571A..16P} cosmology for the VAGC and MPA/JHU stellar masses respectively,
      and finally the multiplicative correction factor used to estimate the mean halo mass for
      central galaxies from the best-fitting NFW mass.}
    \begin{tabular}{cccccc}
\hline\hline
$\log_{10}\left(\frac{M_{*,\mathrm{eff}}}{M_\odot}\right)$ & $\log_{10}\left(\frac{M_{*,\mathrm{eff}}^{\text{MPA}}}{M_\odot}\right)$ & $\log_{10}\left(\frac{\langle M_{200m}\rangle}{h^{-1}M_\odot}\right)$ &$\frac{M_{*,\text{cen}}/M_{200m}}{\Omega_b/\Omega_m}$ &$\frac{M_{*,\text{cen}}^\text{MPA}/M_{200m}}{\Omega_b/\Omega_m}$ &$\frac{M_{200m}}{M_{200m}^{(\text{uncorr})}}$\\
\hline
\multicolumn{6}{c}{Red} \\
10.28 & 10.39 & 12.17$^{+0.19}_{-0.24}$ & 0.056$^{+0.042}_{-0.020}$ & 0.072$^{+0.054}_{-0.025}$ & 1.22 \\
10.58 & 10.70 & 12.14$^{+0.12}_{-0.14}$ & 0.121$^{+0.048}_{-0.030}$ & 0.158$^{+0.062}_{-0.039}$ & 1.36 \\
10.86 & 10.97 & 12.50$^{+0.04}_{-0.05}$ & 0.100$^{+0.011}_{-0.010}$ & 0.129$^{+0.014}_{-0.013}$ & 1.38 \\
11.10 & 11.20 & 12.89$^{+0.04}_{-0.04}$ & 0.071$^{+0.007}_{-0.006}$ & 0.090$^{+0.009}_{-0.008}$ & 1.37 \\
11.29 & 11.38 & 13.25$^{+0.03}_{-0.03}$ & 0.047$^{+0.004}_{-0.003}$ & 0.058$^{+0.004}_{-0.003}$ & 1.31 \\
11.48 & 11.56 & 13.63$^{+0.03}_{-0.03}$ & 0.031$^{+0.002}_{-0.002}$ & 0.037$^{+0.003}_{-0.002}$ & 1.16 \\
11.68 & 11.75 & 14.05$^{+0.05}_{-0.05}$ & 0.019$^{+0.002}_{-0.002}$ & 0.022$^{+0.003}_{-0.002}$ & 1.06 \\
\multicolumn{6}{c}{Blue} \\
10.24 & 10.29 & 11.80$^{+0.16}_{-0.20}$ & 0.120$^{+0.070}_{-0.037}$ & 0.134$^{+0.078}_{-0.041}$ & 1.09 \\
10.56 & 10.63 & 11.73$^{+0.13}_{-0.17}$ & 0.297$^{+0.140}_{-0.076}$ & 0.351$^{+0.165}_{-0.089}$ & 1.11 \\
10.85 & 10.94 & 12.15$^{+0.08}_{-0.10}$ & 0.218$^{+0.054}_{-0.037}$ & 0.265$^{+0.065}_{-0.046}$ & 1.18 \\
11.10 & 11.18 & 12.61$^{+0.10}_{-0.11}$ & 0.136$^{+0.038}_{-0.028}$ & 0.164$^{+0.045}_{-0.034}$ & 1.29 \\
11.28 & 11.35 & 12.69$^{+0.19}_{-0.25}$ & 0.169$^{+0.128}_{-0.059}$ & 0.200$^{+0.152}_{-0.070}$ & 1.45 \\
11.47 & 11.54 & 12.79$^{+0.43}_{-1.01}$ & 0.206$^{+1.908}_{-0.129}$ & 0.243$^{+2.247}_{-0.152}$ & 1.42 \\
11.68 & 11.69 & 12.79$^{+0.58}_{-2.23}$ & 0.338$^{+57.069}_{-0.250}$ & 0.346$^{+58.399}_{-0.256}$ & 1.57 \\
\hline
\end{tabular}

  \end{center}
\end{table*}

\section{Age-matching comparison in detail}\label{app:age-match}

\revall{In this appendix, we present the results of comparing $\Delta\Sigma$ profiles in the $z=0$
  catalog provided by \cite{2014MNRAS.444..729H} with the observed $\Delta\Sigma$ profiles for red
  and blue LBGs and Main sample galaxies in this work.  This comparison avoids the mass inference
  procedure, and depends simply on replicating the LBG selection in those mock catalogs.  \revfinal{While we
  use the $z=0$ simulation snapshot, we weight the LBGs to reproduce the lensing-weighted stellar mass distribution in the
  observed (flux-limited) LBG sample.}}

\begin{figure*}
\begin{center}
\includegraphics[width=0.9\textwidth,angle=0]{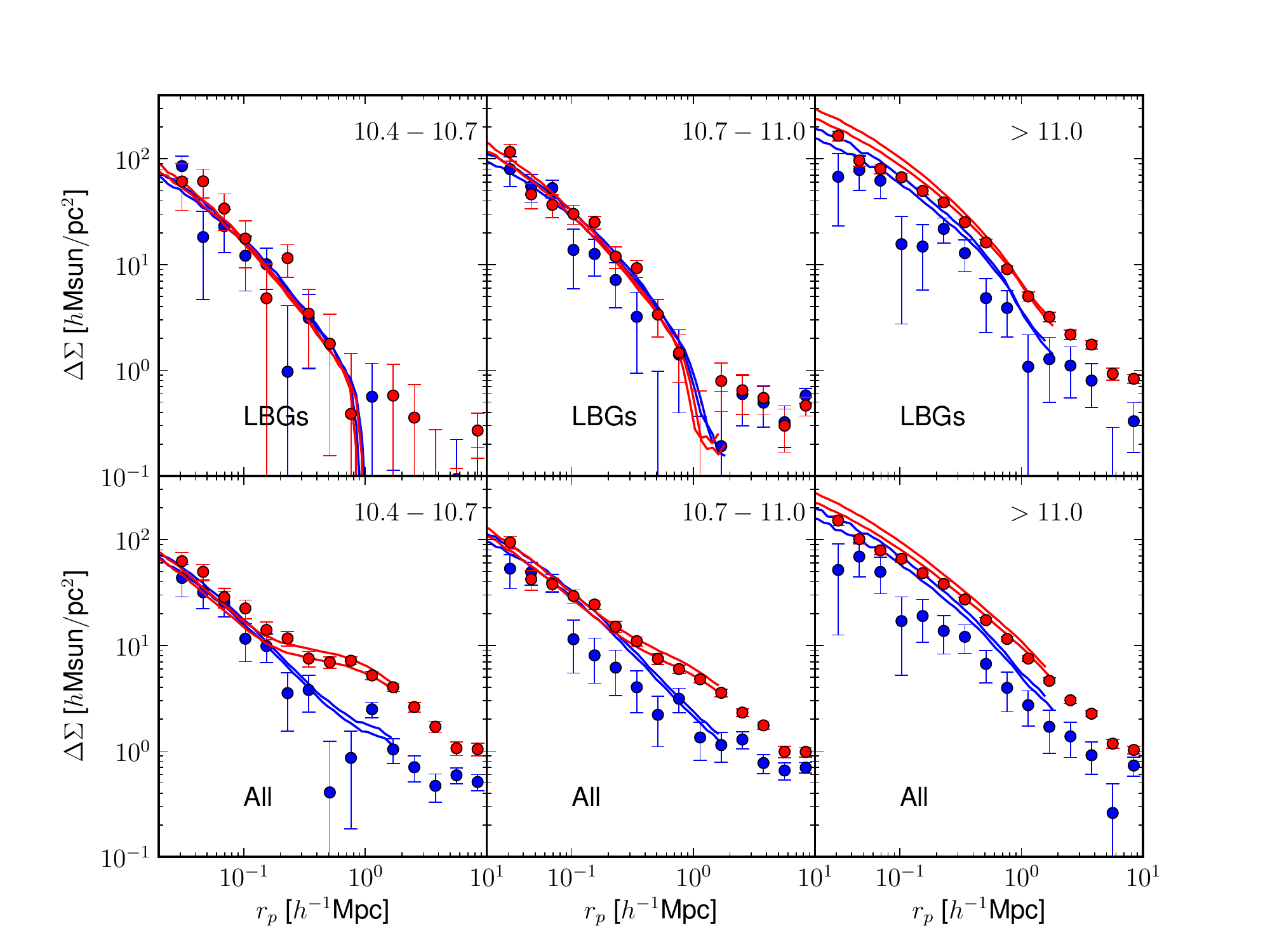}
\caption{\label{fig:age-match} 
\revall{
Points with errorbars show lensing signals for red and blue LBGs (top) and Main sample galaxies without LBG
selection (bottom) in stellar mass bins. The lines show the predictions from the
\protect\cite{2014MNRAS.444..729H} age-matching mock catalogs, with two lines for each sample
(depending on whether we treat their predictions as being fixed in physical or comoving coordinates,
with the latter always being higher than the former).}}
\end{center}
\end{figure*}

\revall{The results of the $\Delta\Sigma$ comparison are shown in Fig.~\ref{fig:age-match}. 
Since our samples are not at $z=0$, we must make a choice as to whether we take the $\Delta\Sigma$ profiles in their $z=0$
simulation as being in physical separations, so no conversion is done to account for the differences
in redshift; or whether their predictions are in comoving coordinates, requiring a conversion to
compare with our results. We show both options in this figure, but the difference between them is
not sufficient to substantially change our conclusions.}  

\revall{\revfinal{As
  shown, for red galaxies, the age-matching predictions are only mildly discrepant with the data.  At
  high mass, the mocks predict significantly higher concentrations for
  red galaxies than are suggested by 
  the observed $\Delta\Sigma$ profiles in the real data.  
The discrepancies between the model and data are particularly evident for the higher mass blue
galaxies (middle and right panels), similar to the results shown in
  Fig.~\ref{fig:compare} when comparing masses.  It is interesting to note that for
  $M_*<10^{11}M_\odot$, the blue masses are higher than the red ones in Fig.~\ref{fig:compare}, but
  this is not strongly reflected in the $\Delta\Sigma$ profiles for LBGs.  This is primarily due to the fact that the model also
  includes a segregation of the red and blue into halos of different concentration, resulting in
  profile differences that partly counteract the effect of the mass differences and result in the
  red profiles being above the blue for  $r_p\lesssim 0.2$\hmpc\ (and the opposite is true for
  larger $r_p$).  For $M_*>10^{11}M_\odot$, Fig.~\ref{fig:compare} shows that the red galaxies have
  higher halo masses than the blue ones, and this is reflected in the $\Delta\Sigma$ profiles;
  however, the $\Delta\Sigma$ profiles for the blue LBGs (or all galaxies) are well above the data
  in this bin.}}

\end{document}